\documentclass[twocolumn,aps,prd,superscriptaddress,nofootinbib,floatfix,10pt]{revtex4-2}

\usepackage[T1]{fontenc}
\usepackage{lmodern}
\usepackage[utf8]{inputenc}
\usepackage[normalem]{ulem}
\usepackage{amsmath,amssymb,bm,mathtools}
\usepackage{euscript}
\usepackage{mathrsfs}
\usepackage{graphicx}

\usepackage[caption=false]{subfig}
\usepackage{tikz}
\usetikzlibrary{arrows.meta}
\usepackage{needspace}
\newsavebox{\triptychbox}
\newlength{\triptychwidth}
\newcommand{\thirdimage}[3]{%
	\begingroup
	\edef\panelcommand{\noexpand\includegraphics[trim=#2 0pt #3 0pt,clip,width=.325\noexpand\textwidth]{#1}}%
	\panelcommand
	\endgroup
}
\newcommand{\threepanels}[2]{%
	\begingroup
	\sbox{\triptychbox}{\includegraphics{#1}}%
	\setlength{\triptychwidth}{\wd\triptychbox}%
	\subfloat[\label{#2:a}]{\thirdimage{#1}{0pt}{\the\dimexpr2\triptychwidth/3\relax}}\hfill
	\subfloat[\label{#2:b}]{\thirdimage{#1}{\the\dimexpr\triptychwidth/3\relax}{\the\dimexpr\triptychwidth/3\relax}}\hfill
	\subfloat[\label{#2:c}]{\thirdimage{#1}{\the\dimexpr2\triptychwidth/3\relax}{0pt}}%
	\endgroup
}

\usepackage{placeins}
\usepackage{xcolor}
\usepackage{hyperref}
\usepackage{booktabs}
\usepackage{multirow}
\usepackage{braket}
\usepackage{physics}
\usepackage{slashed}
\graphicspath{{figures/}}
\allowdisplaybreaks
\newcommand{\CP}{\mathit{CP}}
\newcommand{\Jvec}{\bm J}
\newcommand{\Gmat}{\bm G}

\def \Det{\textmd{Det}}

\def \ie{\textit{i.e.}}
\newbox{\ORCIDicon}
\sbox{\ORCIDicon}{\includegraphics[width=0.8em]{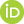}}

\hypersetup{
	pdftitle={Thermal CP dynamics in a four-fermion model: mean-field and functional renormalization group analysis},
	pdfauthor={Arthur F. Vieira and Rudnei O. Ramos},
	colorlinks=true,
	citecolor=blue,
	linkcolor=blue,
	urlcolor=blue
}
\begin{document}
	
	\title{Thermal $\CP$ dynamics in a four-fermion model:\\mean-field and functional renormalization group analysis}
	
	\author{Arthur F. Vieira\,\href{https://orcid.org/0000-0003-2897-2437}{\usebox{\ORCIDicon}}}
	\email{arthur.fvieira@gmail.com}
	\affiliation{Universidade do Estado do Rio de Janeiro, Instituto de F\'isica,
		Departamento de F\'isica Te\'orica, 
		20550-013 Rio de Janeiro, RJ, Brazil}
	
	\author{Rudnei O. Ramos\,\href{https://orcid.org/0000-0001-8781-4928}{\usebox{\ORCIDicon}}}
	\email{rudnei@uerj.br}
	\affiliation{Universidade do Estado do Rio de Janeiro, Instituto de F\'isica,
		Departamento de F\'isica Te\'orica, 
		20550-013 Rio de Janeiro, RJ, Brazil}
	
	\begin{abstract}
		
		\begingroup
		We investigate a one-flavor four-fermion model with scalar, pseudoscalar, and mixed scalar-pseudoscalar interactions at finite temperature and vanishing chemical potential. Our analysis combines the mean-field approximation, which determines the equilibrium condensates, with the functional renormalization group (FRG), which describes the scale dependence of the interactions and the fermion mass phase. Within the mean-field approach, we find that, in the absence of explicit symmetry-breaking sources, the dominant interaction determines the direction of condensation. In particular, a pseudoscalar condensate spontaneously breaks the ordinary $P/\CP$ symmetry and disappears through a continuous thermal phase transition. For regular high-temperature FRG trajectories, the dimensionless mixed coupling tends to zero in the infrared, while its dimensionful counterpart remains finite. Increasing temperature also suppresses the scale evolution of the fermion mass phase. Together, these results clarify the distinct roles of thermal fluctuations in suppressing spontaneous symmetry breaking and fermionic instabilities, without necessarily eliminating explicit symmetry breaking effects.
		\endgroup
		
	\end{abstract}
	
	\maketitle
	
	\section{Introduction}
	\label{sec:introduction}
	
	\begingroup
	The smallness of strong $\CP$ violation remains an open problem
	in quantum chromodynamics (QCD)~\cite{Abel:2020pzs,Liang:2023jfj}.
	The most general renormalizable QCD Lagrangian density contains the
	topological term
	\begin{equation}
		\mathcal L_{\theta}=
		\frac{g_s^2\theta}{32\pi^2}
		G^a_{\mu\nu}\tilde G^{a\mu\nu},
		\label{eq:theta-term}
	\end{equation}
	where $g_s$ is the $SU(3)$ gauge coupling, $a=1,\ldots,8$ is the
	adjoint color index, and $G^a_{\mu\nu}$ is the non-Abelian field-strength
	tensor.  We use $\epsilon^{0123}=+1$ to define its dual:
	$\tilde G^{a\mu\nu}\equiv
	\epsilon^{\mu\nu\rho\sigma}G^a_{\rho\sigma}/2$.  The angle $\theta$ is the QCD vacuum angle.
	An anomalous axial transformation of the quark fields moves this phase into
	the generally complex quark mass matrix $M_q$, which acts in flavor space.
	The observable, basis-independent parameter is consequently
	\begin{equation}
		\bar\theta=\theta+\arg\det M_q.
		\label{eq:theta-bar}
	\end{equation}
	The neutron electric dipole moment (nEDM) limit, $ |d_n|<1.8\times10^{-26}\,e\,\mathrm{cm}$,
	combined with the lattice-QCD estimate, $d_n\simeq-1.5\times10^{-3}\,\bar\theta\,e\,\mathrm{fm}$, yields the inferred bound
	$|\bar\theta|\lesssim1.2\times10^{-10}$~\cite{Abel:2020pzs,Liang:2023jfj}.
	Explaining why this angle is so small is the strong-$\CP$ problem.
	\endgroup
	The Peccei-Quinn mechanism dynamically relaxes $\bar\theta$ by promoting it to a field degree of freedom associated with a new anomalous axial symmetry~\cite{Peccei:1977hh,Weinberg:1977ma,Wilczek:1977pj}.  An alternative class of proposals assumes that $P$ or $\CP$ is exact in the microscopic	action and is broken only by the vacuum.  In this case, the effective potential develops degenerate minima related by the discrete symmetry, and the choice of a minimum yields a $P$- and $\CP$-odd expectation value \cite{Nelson:1983zb,Barr:1984qx,Barr:1984fh,Babu:1989rb,Barr:1991qx,
		Mohapatra:1978fy,Beg:1978mt,Georgi:1978xz,Dine:1993qm,Dine:2015jga,Csaki:2024lvk}.
	
	At finite temperature, condensate suppression must be distinguished
	from symmetry restoration~\cite{Klevansky:1992qe,Hatsuda:1994pi,Buballa:2003qv,Guan:2026tbd}. Thermal occupation factors can restore a spontaneous vacuum-broken symmetry. However, they do not remove an explicit breaking term from the microscopic action. A condensate induced by such a source may decrease smoothly with temperature, defining a crossover rather than a symmetry-restoring transition.

	Nambu-Jona-Lasinio (NJL) and related four-fermion models provide a simple setting for this distinction~\cite{Klevansky:1992qe,Hatsuda:1994pi,Buballa:2003qv,Braun:2011pp}. Their local interactions can arise in low-energy truncations of QCD, and attractive scalar or pseudoscalar channels can generate a fermion mass dynamically. Although these models do not describe confinement, they allow a direct study of symmetry breaking and thermodynamics. Strong infrared growth of a four-fermion vertex signals a condensate instability~\cite{Braun:2011pp,Braun:2017srn}, while a Hubbard-Stratonovich transformation introduces auxiliary fields for the corresponding collective channels.
	
	The model studied here is the finite-temperature extension of the
	$\CP$-violating four-fermion sector analyzed at zero temperature
	\cite{Huang:2024ypj}.  Its distinctive mixed interaction is
	\begin{equation}
		\mathcal O_{SP}
		= (\bar\psi\psi)(\bar\psi i\gamma_5\psi).
		\label{eq:OSP-introduction}
	\end{equation}
	
	The scalar bilinear is $P$ even and $C$ even, whereas the pseudoscalar is $P$ odd and $C$ even. Their product is therefore odd under ordinary $P/\CP$ and couples the two gap equations. Its coefficient also breaks the continuous $U(1)_{\rm A}$ symmetry. These assignments refer to a specified {scalar-pseudoscalar} basis. {The vacuum analysis of this mixed channel have been studied in a four-fermion model~\cite{Huang:2024ypj} and its dynamical generation has also been demonstrated with gauge interactions~\cite{Guan:2026tbd}.}
	
	At mean-field level, homogeneous auxiliary fields determine the fermion determinant and gap equations at leading order in a large number of colors ($N_c$). This approximation does not determine the scale dependence of the interaction vertices or include fluctuations of the collective fields. Moreover, a restricted set of four-fermion channels is generally not closed under renormalization, since other symmetry-allowed operators can feed back on the retained interactions~\cite{Berges:2000ew,Braun:2011pp,Braun:2017srn}.
	
	Comparisons beyond the mean-field, therefore, require consistent channels and parameters. For example, optimized perturbation theory includes finite-$N_c$ corrections in the two-flavor NJL model~\cite{Kneur:2010yv}, while finite-$N$ corrections can modify thermal scales and produce a tricritical point in the planar Gross-Neveu model~\cite{Kneur:2007vm}. Related planar Gross-Neveu and Thirring calculations also illustrate the sensitivity of finite-density results to channel consistency~\cite{Martins:2024dag}. These examples motivate such checks, but do not supply corrections to the four-dimensional one-flavor model studied here.
	
	Here we use the FRG to follow the coupled interactions from an ultraviolet (UV) scale to the infrared (IR)~\cite{Wetterich:1992yh,Berges:2000ew,Braun:2011pp}. Fermionic Matsubara modes enter the finite-temperature threshold functions and suppress the loop contributions as $T/k$ increases, for a given temperature $T$ and scale $k$. Our leading-$N_c$ pointlike truncation resums fermionic vertex corrections in scalar, pseudoscalar, and mixed channels~\cite{Braun:2011pp,Huang:2024ypj}. It complements the equilibrium mean-field calculation but does not include mesonic fluctuation corrections to its phase diagram.
	
	Thermal $\CP$ breaking and restoration have been investigated in multiflavor NJL models, with particular attention to the effects of finite density \cite{Chatterjee:2011yz}, the electromagnetic scale anomaly \cite{YuanyuanWang:2025cim}, and the implications of $\CP$-odd backgrounds for topological susceptibility and axionic domain walls \cite{Huang:2024nbd}. Related effects of $\CP$ phases have also been explored in the context of hybrid-star matter \cite{Lopes:2022efy}.
	
	{Our purpose is to examine how finite temperature changes this coupled scalar-pseudoscalar sector in a one-flavor theory at vanishing chemical potential and external field. We follow the running interactions and mass phase, and use mean-field thermodynamics to distinguish spontaneous symmetry restoration from the thermal decrease of a condensate induced by an explicit source.}
	
	To isolate these mechanisms, we use a one-flavor theory with $N_c=3$ and benchmark parameters rather than a fit to hadronic observables. {\color{black}The thermal scales characterize this effective model and serve as qualitative guides to the dynamics.} A phenomenological application would require an enlarged flavor structure and a consistent calibration.
	
	The paper is organized as follows. Section~\ref{sec:general-construction} introduces the model and its symmetry classification. We derive the mean-field potential in Sec.~\ref{sec:meanfield}, then discuss source-free condensation and explicit-source responses in Secs.~\ref{sec:critical} and~\ref{sec:results}. Sections~\ref{sec:FRGsetup} and~\ref{sec:frg-results} present the finite-temperature FRG equations and results, including their connection with mean-field thermodynamics. We conclude in Sec.~\ref{sec:Conclusions}. {\color{black}Three appendices provide details of the derivations.}

	\Needspace{8\baselineskip}
	\section{QCD-inspired model\\and symmetries}
	\label{sec:general-construction}
	
	{Let us start by first introducing the QCD-inspired model and the symmetry criteria used throughout our analysis.}
	
	\subsection{Model and conventions}
	
	The QCD motivation fixes the phase convention used in the effective
	model. Consider $N_f$ degenerate quark flavors in the fundamental
	representation of $SU(N_c)$, with the UV Lagrangian density
	\begin{align}
		\mathcal{L}_{\rm UV}=&-\frac14G^a_{\mu\nu}G^{a\mu\nu}
		+\bar\psi\left(i\gamma^\mu D_\mu-m e^{i\gamma_5\theta_m/2}\right)\psi
		\nonumber\\
		&+\frac{g_s^2\theta_g}{32\pi^2}
		G^a_{\mu\nu}\tilde G^{a\mu\nu},
		\label{eq:general-UV-action}
	\end{align}
    where 
		$G^a_{\mu\nu}=\partial_\mu A^a_\nu-\partial_\nu A^a_\mu
		-g_sf^{abc}A^b_\mu A^c_\nu$ and
		$f^{abc}$ are
		the $SU(N_c)$ structure constants.
	Here, $\theta_g$ multiplies the topological term and $\theta_m/2$
	is the common mass phase. The covariant derivative is
	$D_\mu=\partial_\mu+i g_s A_\mu^aT^a$, with $T^a$ the generators of the group.
	The axial field redefinition
	\begin{equation}
		\psi\longmapsto\psi'=e^{i\alpha\gamma_5}\psi,
		\qquad
		\bar\psi\longmapsto\bar\psi'=\bar\psi e^{i\alpha\gamma_5},
		\label{eq:general-axial-rotation}
	\end{equation}
	changes the mass phase and, through the anomalous fermion measure,
	the topological angle~\cite{Fujikawa:1979ay,bertlmann2000anomalies}:
	\begin{equation}
		\theta_m\longmapsto\theta_m+4\alpha,
		\qquad
		\theta_g\longmapsto\theta_g-2N_f\alpha.
		\label{eq:general-theta-shifts}
	\end{equation}
	Thus, the combination
	\begin{equation}
		\theta\equiv\theta_m+\frac{2\theta_g}{N_f}
		\label{eq:general-physical-phase}
	\end{equation}
	{is unchanged and satisfies $\theta=2\bar\theta/N_f$ in the convention of Sec.~\ref{sec:introduction}.} Choosing $\alpha=\theta_g/(2N_f)$
	places this phase entirely in the mass term:
	\begin{equation}
		\mathcal{L}_{\rm UV}'=-\frac14G^a_{\mu\nu}G^{a\mu\nu}
		+\bar\psi\left(i\gamma^\mu D_\mu-m e^{i\gamma_5\theta/2}\right)\psi.
		\label{eq:general-mass-phase-action}
	\end{equation}
	
	We now use this mass-phase convention in a local four-fermion
	description, with the gauge fields integrated out. The effective
	model retains the scalar, pseudoscalar, and mixed interactions:
	\begin{align}
		\mathcal{L}_{\rm 4f}=&
		\bar\psi\left(i\gamma^\mu\partial_\mu
		-m e^{i\gamma_5\theta/2}\right)\psi\nonumber\\
		&+{\frac{G_S}{2}(\bar\psi\psi)^2
			+\frac{G_P}{2}(\bar\psi i\gamma_5\psi)^2}\nonumber\\
		&+{C_4(\bar\psi\psi)(\bar\psi i\gamma_5\psi)},
		\label{eq:general-four-fermion-action}
	\end{align}
	{The mixed term couples the scalar and pseudoscalar responses and is odd
		under $P$ and $\CP$.} {We retain these channels because they describe the
		homogeneous condensates studied below, while additional four-fermion
		operators are omitted.}\footnote{The retained channels do not
		form a Fierz-complete basis. Further $P$-odd operators include
		the tensor channel
		$(\bar\psi\sigma_{\mu\nu}\psi)
		(\bar\psi i\sigma^{\mu\nu}\gamma_5\psi)$ and the color-exchange
		channels $(\bar\psi_i\psi_j)(\bar\psi_j i\gamma_5\psi_i)$ and
		$(\bar\psi_i\sigma_{\mu\nu}\psi_j)
		(\bar\psi_j i\sigma^{\mu\nu}\gamma_5\psi_i)$, with color indices
		$i,j$. {These operators can mix with the {scalar-pseudoscalar} interaction}
		under RG evolution. A treatment retaining their independent
		channels would require additional projections beyond the scalar
		and pseudoscalar fields used here~\cite{Guan:2026tbd,Huang:2024ypj}.}
	
	The interaction sector preserves continuous $U(1)_{\rm A}$
	symmetry when $G_S=G_P$ and $C_4=0$. Either a difference between
	$G_S$ and $G_P$ or a nonzero $C_4$ breaks that symmetry.
	In the gauge-free path integral, the axial anomaly is not
	recomputed. Instead, $\theta$ is a mass-phase input matched to
	the underlying QCD-like theory. An axial field redefinition
	rotates the mass and the interactions together. It can remove
	the mass phase without changing the four-fermion couplings
	when $G_S=G_P$ and $C_4=0$, but not in general.
	
	For the single-flavor model used in the calculation, the thermal
	Euclidean action is
	\begin{align}
		S_{\rm E}=&\int_x
		\bigg[\bar\psi
		\left(\gamma_\mu\partial_\mu+m_S+i\gamma_5m_P\right)\psi
		\nonumber\\
		&-{\frac{G_S}{2}(\bar\psi\psi)^2}
		-{\frac{G_P}{2}(\bar\psi i\gamma_5\psi)^2}\nonumber\\
		&-{C_4(\bar\psi\psi)(\bar\psi i\gamma_5\psi)}\bigg],
		\label{eq:bare-action}
	\end{align}
	with
	\begin{equation}
		\int_x\equiv\int_0^\beta\dd\tau\int_{\mathbb R^3}\dd^3\vb*{x},
		\qquad \beta=1/T,
	\end{equation}
	and $m_S=m\cos(\theta/2)$, $m_P=m\sin(\theta/2)$.
	Fermions obey antiperiodic boundary conditions in Euclidean
	time. The UV cutoff $\Lambda$ sets the reference scale of
	this effective model and is distinct from
	$\Lambda_{\overline{\rm MS}}$  as used, e.g. in the modified minimal subtraction ($\overline{\textmd{MS}}$) regularization scheme. For convenience, we choose to present all our numerical results in units of $\Lambda$. Hence, for orientation only,
	choosing $\Lambda=1\,\mathrm{GeV}$ converts the quoted
	dimensionless temperatures directly into GeV. This is only
	a choice of reference units, consistent with the benchmark
	scope stated in Sec.~\ref{sec:introduction}.
	
	Since the scalar bilinear is parity even and the pseudoscalar
	bilinear is parity odd, ordinary $P/\CP$ is a symmetry of this action when
	\begin{equation}
		C_4=0,
		\qquad m_P=m\sin(\theta/2)=0.
		\label{eq:exact-symmetry}
	\end{equation}
	A scalar mass preserves parity, whereas $m_P$ and $C_4$
	are explicit parity-odd sources. A nonzero pseudoscalar
	condensate is then a spontaneous order parameter only
	when both sources vanish.
	
	
	\section{Finite-temperature mean-field construction}
	\label{sec:meanfield}
	
	Let us introduce the scalar and pseudoscalar auxiliary fields so
	that the fermions can be integrated out and then use
	their effective potential to obtain the gap equations.  Then, we can compute the stationary points, which correspond to  the equilibrium
	solutions.
	
	\subsection{Hubbard-Stratonovich transformation}
	
	The Hubbard-Stratonovich (HS) transformation replaces the
	four-fermion interactions by couplings to a real scalar field $\sigma$
	and a real pseudoscalar field $\eta$. For the corresponding Gaussian
	integral to converge, the couplings must satisfy
	\begin{equation}
		G_S>0,
		\qquad \Delta_G\equiv G_SG_P-C_4^2>0.
		\label{eq:HS-domain}
	\end{equation}
	{These conditions also imply $G_P>0$ and allow us to write the
		equivalent bosonized action}
	\begin{equation}
		\begin{aligned}
			S_{\rm B}=&\int_x\bigg\{\bar\psi
			(\gamma_\mu\partial_\mu+M_S+i\gamma_5M_P)\psi
			\\
			&+\frac{G_P\sigma^2-2C_4\sigma\eta+G_S\eta^2}
			{2\Delta_G}\bigg\},
		\end{aligned}
		\label{eq:bosonized-action}
	\end{equation}
	where $M_S=m_S+\sigma$ and $M_P=m_P+\eta$ include the
	external sources and the auxiliary fields. {This rewriting is
		exact, as shown by the Gaussian identity in
		Appendix~\ref{app:meanfield-construction}.}
	
	\subsection{Thermodynamic-potential density}
	
	At mean-field, we take the auxiliary fields to be background homogeneous fields,
	$\sigma(x)=\bar\sigma$ and $\eta(x)=\bar\eta$. The mass
	components are $M_S=m_S+\bar\sigma$ and $M_P=m_P+\bar\eta$,
	with magnitude and fermion energy given,
	respectively, by
	\begin{equation}
		M^2=M_S^2+M_P^2,
		\qquad E_p=\sqrt{p^2+M^2}.
		\label{eq:thermal-dispersion}
	\end{equation}
	Integrating out the fermions gives the thermodynamic-potential
	density,
	\begin{equation}
		\begin{aligned}
			\Omega_T(\bar\sigma,\bar\eta)=&
			\frac{G_P\bar\sigma^2+G_S\bar\eta^2-2C_4\bar\sigma\bar\eta}
			{2\Delta_G}
			\\
			&+F_T(M).
		\end{aligned}
		\label{eq:Omega-bar}
	\end{equation}
	The positive quadratic term opposes nonzero mean fields,
	whereas the fermion contribution $F_T$ can favor them. Their
	competition determines whether condensates form and how they
	respond to temperature. With the spatial cutoff $\Lambda_3$, this fermion
	contribution is
	\begin{equation}
		\begin{aligned}
			F_T(M)=&-\frac{N_c}{\pi^2}
			\int_0^{\Lambda_3}\dd p\,p^2
			\bigg[E_p+2T\ln\!\left(1+e^{-E_p/T}\right)\bigg],
		\end{aligned}
		\label{eq:fermion-loop}
	\end{equation}
	where here $p=|\vb* p|$ is the radial three-momentum.
	A field-independent constant is omitted. At zero chemical
	potential, $\Omega_T$ is also the Helmholtz free-energy
	density. The determinant and sum over Matsubara frequencies are
	evaluated in Appendix~\ref{app:meanfield-construction}.
	
	\subsection{Gap equations and condensates}
	\label{sec:gap_equations_condensates}
	
	The gap equations follow by setting the derivatives of
	$\Omega_T$ with respect to both background mean fields to zero.
	The fermion contribution satisfies
	\begin{align}
		\frac{\partial F_T}{\partial M_i}
		&=-\EuScript{K}_3(M,T)M_i,
		\qquad i=S,P,
		\label{eq:fermion-loop-derivative}\\
		\EuScript{K}_3(M,T)
		&=\frac{N_c}{\pi^2}\int_0^{\Lambda_3}\dd p\,
		\frac{p^2}{E_{\vb* p}}
		\tanh\!\left(\frac{E_{\vb* p}}{2T}\right).
		\label{eq:kernel}
	\end{align}
	The kernel $\EuScript{K}_3$ is the common fermion-loop
	factor in both channels \textcolor{black}{and $\Lambda_3$ is the UV cutoff on the magnitude of the three-momentum.} Its analytic properties are
	collected in Appendix~\ref{app:kernel}. The gap equations then obtained are
	\begin{align}
		\frac{\partial\Omega_T}{\partial\bar\sigma}
		&=\frac{G_P\bar\sigma-C_4\bar\eta}{\Delta_G}
		-\EuScript{K}_3(M,T)M_S=0,
		\label{eq:gap-scalar}\\
		\frac{\partial\Omega_T}{\partial\bar\eta}
		&=\frac{G_S\bar\eta-C_4\bar\sigma}{\Delta_G}
		-\EuScript{K}_3(M,T)M_P=0.
		\label{eq:gap-pseudoscalar}
	\end{align}
	The gap equations given above have stationary solutions, with the
	equilibrium one corresponding to that with the lowest
	thermodynamic potential:
	\begin{equation}
		(\bar\sigma_{\rm phys},\bar\eta_{\rm phys})(T)
		\in\operatorname*{arg\,min}_{(\bar\sigma,\bar\eta)}
		\Omega_T(\bar\sigma,\bar\eta).
		\label{eq:global-minimum}
	\end{equation}
	Several degenerate minima are possible, for example when
	they are related by an exact symmetry.
	
	The background mean fields are related to the fermion condensates
	through~\cite{Pannullo:2024hqj,Winstel:2024dqu}
	\begin{align}
		\langle\bar\psi\psi\rangle_T
		&=-\frac{G_P\bar\sigma-C_4\bar\eta}{\Delta_G}
		=-\EuScript{K}_3(M,T)M_S,
		\label{eq:scalar-condensate}\\
		\langle\bar\psi i\gamma_5\psi\rangle_T
		&=-\frac{G_S\bar\eta-C_4\bar\sigma}{\Delta_G}
		=-\EuScript{K}_3(M,T)M_P.
		\label{eq:pseudoscalar-condensate}
	\end{align}
	The underlying HS identity is explained in
	Appendix~\ref{app:meanfield-construction}. Thus, the mean
	fields and condensates are related but are not the same
	quantities. At fixed cutoff and couplings, increasing
	temperature suppresses the mean fields and condensates,
	so that $(M_S,M_P)\to(m_S,m_P)$ at high temperature.
	This limit does not remove the explicit sources $m_P$
	or $C_4$. A short derivation is given in
	Appendix~\ref{app:high-temperature-limit}.
	
	\section{Source-free phase structure\\and thermal restoration}
	\label{sec:critical}
	
	To isolate spontaneous symmetry breaking, we start by first setting  $m_S=m_P=C_4=0$. Although the microscopic mass vanishes, a sufficiently attractive interaction can still generate a nonzero constituent mass $M$.
	
	\subsection{Condensate direction and vacuum instability}
	\label{sec:channel-selection}
	
	The relative interaction strengths determine which channel condenses. To see this, we use $M_S=\bar\sigma$ and $M_P=\bar\eta$ in the source-free limit, for which the potential becomes
	\begin{equation}
		\Omega_T(M_S,M_P)=\frac{M_S^2}{2G_S}+\frac{M_P^2}{2G_P}
		+F_T\!\left(\sqrt{M_S^2+M_P^2}\right).
		\label{eq:sourcefree-potential}
	\end{equation}
	At fixed $M$, the fermion term is independent of the condensate direction. The larger coupling then selects the preferred axis. For $G_P>G_S$, writing $M_S=M\cos\varphi$ and $M_P=M\sin\varphi$ gives
	\begin{equation}
		\begin{aligned}
			& \Omega_T(M,\varphi)-\Omega_T(M,\pi/2)
			\\
			&=\frac{M^2\cos^2\varphi}{2}
			\left(\frac1{G_S}-\frac1{G_P}\right)\geq0.
		\end{aligned}
		\label{eq:pseudoscalar-selection}
	\end{equation}
	{For nonzero $M$, this inequality is strict away from the pseudoscalar axis. Hence, the minimum has $M_S=0$, rather than requiring this condition as an additional assumption.}
	
	An interaction must also be strong enough to destabilize the zero-mass state. {\color{black}We normalize the couplings to the zero-temperature critical coupling, defined by}
	\begin{equation}
		g_c\equiv\frac{1}{\EuScript{K}_3(0,0)}=\frac{4\pi^2}{N_c \Lambda^2},
		\qquad g_c\Lambda^2=\frac{4\pi^2}{N_c}\simeq13.2,
		\label{eq:gc}
	\end{equation}
	where we have considered $N_c=3$ and $\Lambda_3/\Lambda=1/\sqrt2$, which will be used throughout the benchmark values below. The cutoff convention and the expansion establishing this \textcolor{black}{critical coupling} are given in Appendix~\ref{app:kernel}. A decoupled channel forms a nonzero vacuum condensate when its coupling exceeds $g_c$.
	
	\begin{figure}[t]
		\centering
		\includegraphics[width=0.93\columnwidth]{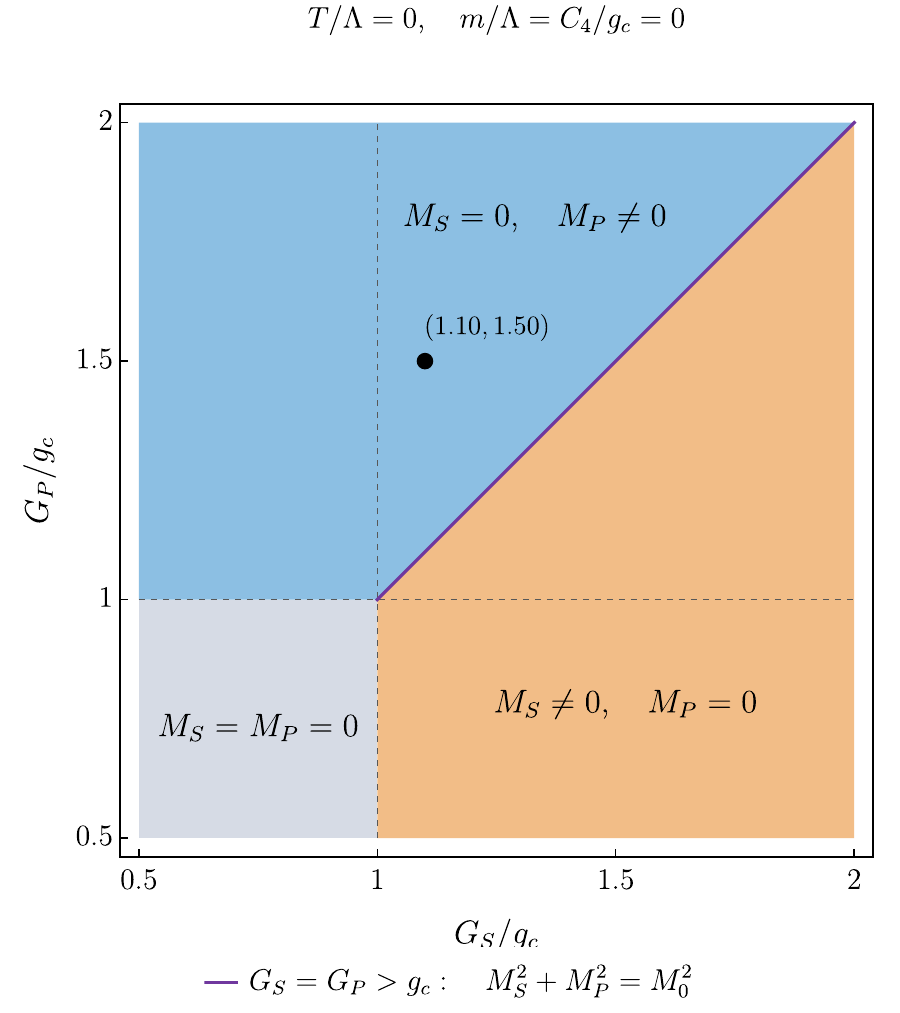}
		\caption{Source-free zero-temperature phase selection in the
			$(G_S/g_c,G_P/g_c)$ plane. On $G_S=G_P>g_c$, the axial orientation is degenerate.
			The black point marks the pseudoscalar benchmark
			$(G_S/g_c,G_P/g_c)=(1.10,1.50)$.}
		\label{fig:zeroT-phase-map}
	\end{figure}
	
	Figure~\ref{fig:zeroT-phase-map} combines the strength and direction criteria. The figure illustrates that when both couplings lie below $g_c$, the equilibrium mass vanishes. Otherwise, the stronger channel selects the condensate direction. When $G_S=G_P>g_c$, neither direction is preferred, and the axial orientation is degenerate.
	
	For the pseudoscalar benchmark we choose
	\begin{equation}
		G_S=1.10g_c,\qquad G_P=1.50g_c,\qquad m=C_4=0.
		\label{eq:CP-benchmark}
	\end{equation}
	The two equilibrium branches satisfy
	\begin{equation}
		M_S=0,\qquad M_P=\pm M_0(T),\qquad
		\EuScript{K}_3(M_0(T),T)=G_P^{-1}.
		\label{eq:pseudoscalar-gap}
	\end{equation}
	Their condensates are given by
	\begin{equation}
		\langle\bar\psi\psi\rangle_T=0,\qquad
		\langle\bar\psi i\gamma_5\psi\rangle_T=\mp M_0(T)/G_P.
		\label{eq:broken-condensates}
	\end{equation}
	The opposite signs label parity related minima. Choosing either minimum, therefore, spontaneously breaks the ordinary $P/\CP$.
	
	\begin{figure*}[!t]
		\centering
		\threepanels{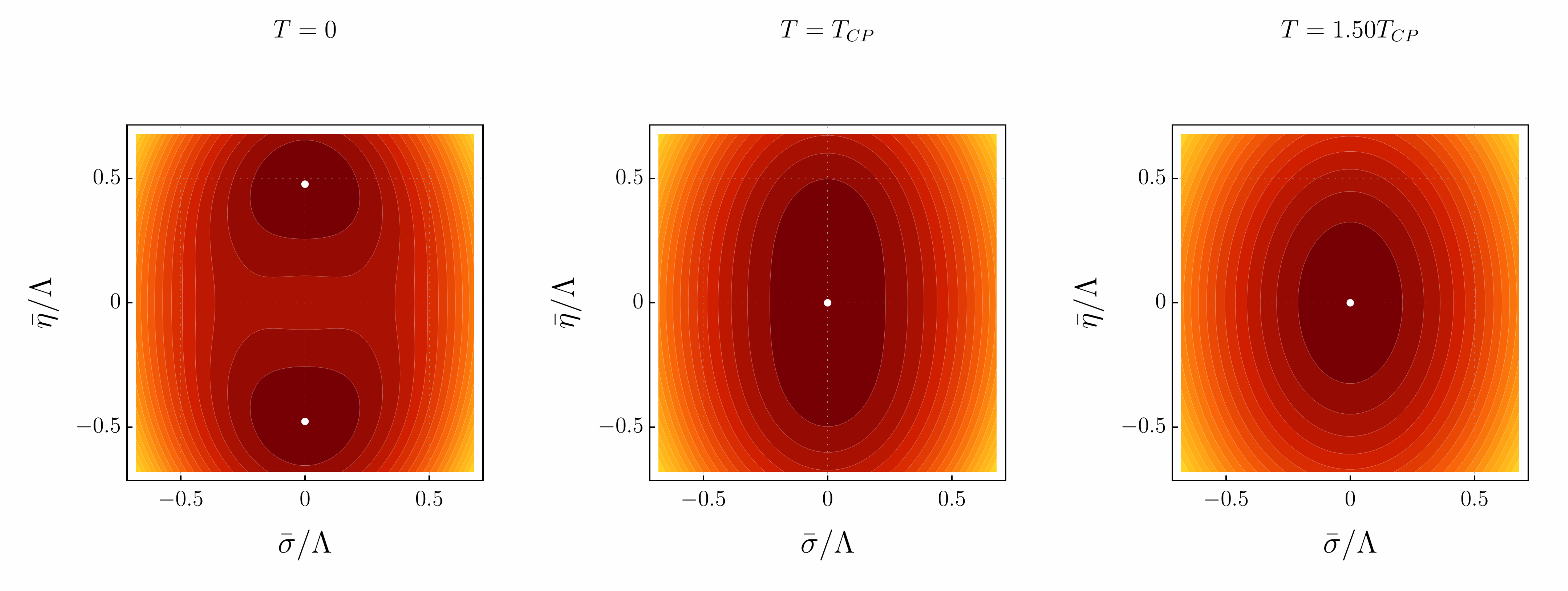}{fig:CP-contours}
		\caption{Subtracted thermodynamic-potential density
			$\Delta\Omega_T(\bar\sigma,\bar\eta)$ for $G_S=1.10g_c$,
			$G_P=1.50g_c$, $m=C_4=0$, $N_c=3$, and
			$\Lambda_3/\Lambda=1/\sqrt2$.
			The axes $\bar\sigma/\Lambda$ and $\bar\eta/\Lambda$ coincide respectively with $(M_S/\Lambda,M_P/\Lambda)$ because $m=0$. White points mark global minima. Darker
			colors denote lower potential values. }
		\label{fig:CP-contours}
	\end{figure*}
	
	\subsection{Thermal restoration\\of spontaneous \texorpdfstring{$P/\CP$}{P/CP} breaking}
	\label{sec:critical-temperatures}
	
\begin{figure}[t]
		\centering
		\includegraphics[width=0.90\columnwidth]{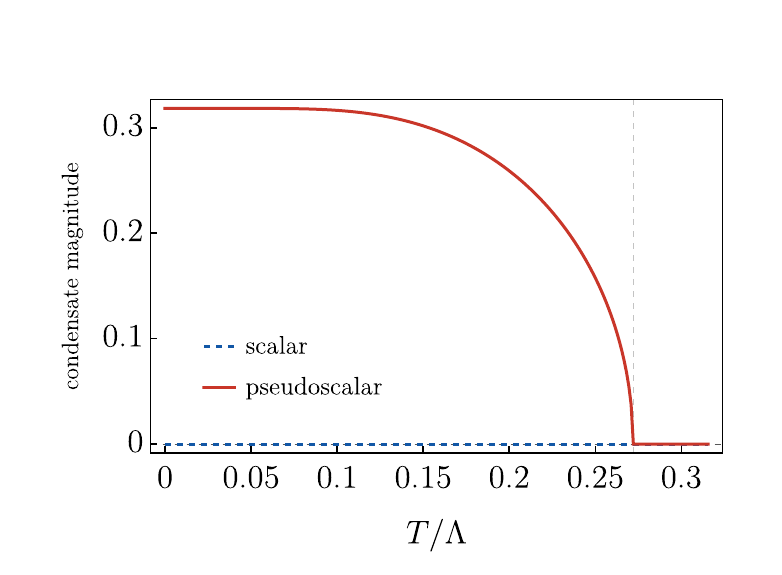}
		\caption{Temperature dependence of the condensates for spontaneous
			$P/\CP$ breaking and thermal restoration with
			$G_S=1.10g_c$, $G_P=1.50g_c$, $m=C_4=0$.
			The blue dashed scalar and red solid pseudoscalar curves show,
			respectively, $g_c|\langle\bar\psi\psi\rangle_T|/\Lambda$ and
			$g_c|\langle\bar\psi i\gamma_5\psi\rangle_T|/\Lambda$.
			The vertical line marks $T_{\CP}/\Lambda=0.272$, where the
			pseudoscalar order parameter vanishes continuously.}
		\label{fig:CP-restoration}
	\end{figure}

	Figure~\ref{fig:CP-contours} illustrates how the two minima approach each other and merge at the origin as the system is heated. Thus, the pseudoscalar condensate decreases continuously to zero, as shown in Fig.~\ref{fig:CP-restoration}. {\color{black}To display the potential landscapes, we subtract the equilibrium value at each temperature, defining}
	\begin{equation}
		{\color{black}\begin{aligned}
				\Delta\Omega_T(\bar\sigma,\bar\eta)
				&=\Omega_T(\bar\sigma,\bar\eta)\\
				&\quad-\Omega_T\big(\bar\sigma_{\rm eq}(T),\bar\eta_{\rm eq}(T)\big)
				\geq0.
		\end{aligned}}
		\label{eq:subtracted-potential}
	\end{equation}
	{\color{black}Here, $\bar\sigma_{\rm eq}(T)$ and $\bar\eta_{\rm eq}(T)$ locate a global minimum of the potential.}
	Near the transition, the potential along the pseudoscalar axis has the following form when expanded like in a Ginzburg-Landau form, 
	\begin{equation}
		\begin{aligned}
			\Omega_T(0,M_P)-\Omega_T(0,0)
			=&\frac12a(T)M_P^2+\frac14b(T)M_P^4
			\\ &+O(M_P^6),
		\end{aligned}
		\label{eq:landau-expansion}
	\end{equation}
	with $a(T)=G_P^{-1}-\EuScript{K}_3(0,T)$. The quartic coefficient is positive, as shown in Appendix~\ref{app:kernel}. Hence, the nonzero minimum approaches the origin continuously as $a(T)$ reaches zero. The restoration temperature is fixed by
	\begin{equation}
		\EuScript{K}_3(0,T_{\CP})=G_P^{-1}.
		\label{eq:TCP-onset}
	\end{equation}
	Using the values given in Eq.~\eqref{eq:CP-benchmark}, we then obtain $T_{\CP}/\Lambda=0.272$ and $M_0(0)/\Lambda=0.478$.
	
	
	Above $T_{\CP}$, the kernel decreases with $M$ and is already smaller than $G_P^{-1}$ at $M=0$. This means that Eq.~\eqref{eq:pseudoscalar-gap} has no nonzero solution. Together with the positive quartic coefficient, this establishes a second-order restoration of spontaneously broken $P/\CP$ within the mean-field theory, rather than an exchange of stability between separated minima.
	The pseudoscalar orientation reflects the chosen ordering $G_P>G_S$. Reversing this ordering, with $G_S>g_c$, instead selects a scalar condensate. Below we also consider a case in which both components are nonzero.
	
	\Needspace{6\baselineskip}
	\section{Explicit sources\\and thermal response}
	\label{sec:results}
	
	{We next include a mass phase or a mixed interaction, which can select a preferred condensate direction. A particularly interesting case is a purely pseudoscalar mass. Although the mass is nonzero, the scalar condensate can still vanish through a genuine thermal transition that restores a remaining reflection symmetry. We first compare the responses for different mass phases, then explain this transition in Sec.~\ref{sec:residual-transition} and examine the effect of the mixed interaction.}
	
	\begin{figure*}[!t]
		\centering
		\threepanels{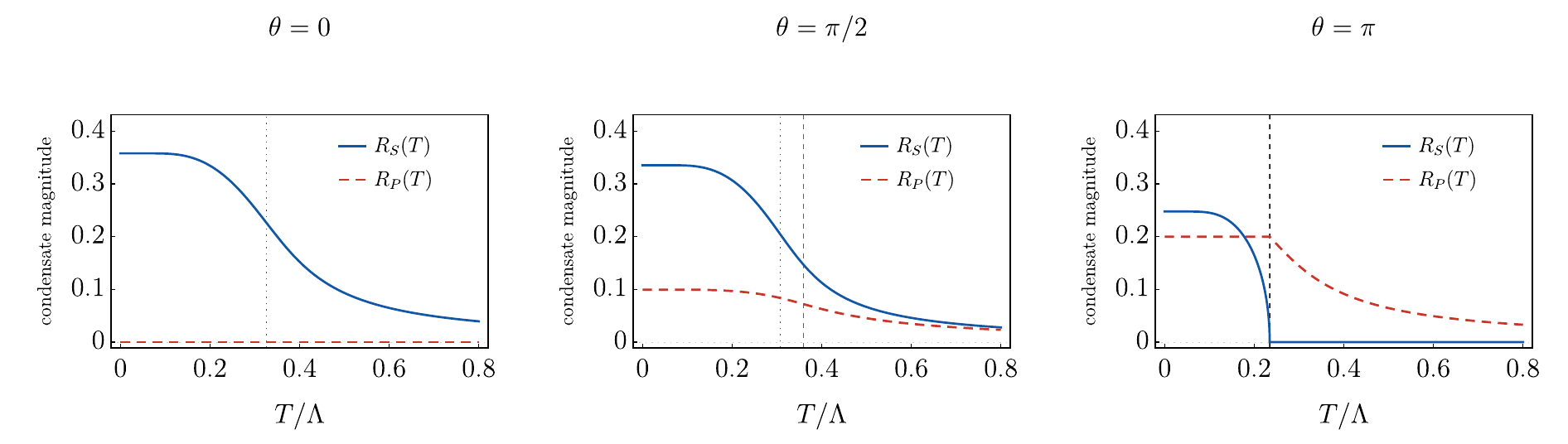}{fig:mass-phase}
		\caption{Scalar (blue solid) and pseudoscalar (red dashed) condensate magnitudes $R_S$ and $R_P$ for $G_S=1.50g_c$, $G_P=1.10g_c$, $C_4=0$, $m/\Lambda=0.08$. Blue dotted and red dashed vertical lines mark $T_{\rm pc}^{(S)}$ and $T_{\rm pc}^{(P)}$, respectively. At $\theta=0$, $T_{\rm pc}^{(S)}/\Lambda\simeq0.326$, while $R_P=0$ and no pseudoscalar crossover scale is assigned. At $\theta=\pi/2$, the scales are $T_{\rm pc}^{(S)}/\Lambda\simeq0.307$ and $T_{\rm pc}^{(P)}/\Lambda\simeq0.359$. At $\theta=\pi$, the black dashed line marks the genuine scalar-reflection transition at $T_{\rm r}/\Lambda\simeq0.235$. Numerical scales are collected in Table~\ref{tab:thermal-scales}.}
		\label{fig:mass-phase}
	\end{figure*}
	
	\begin{figure*}[!t]
		\centering
		\threepanels{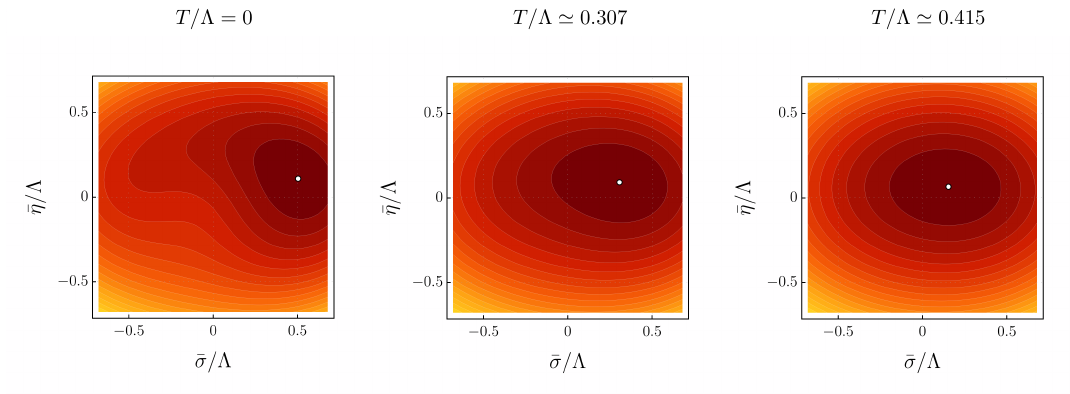}{fig:mass-contours}
		\caption{Potential contour levels in the plane $(\bar \sigma, \bar \eta)$ for different values of temperature and for the explicit-mass benchmark
			$G_S=1.50g_c$, $G_P=1.10g_c$, $C_4=0$, $m/\Lambda=0.08$,
			$\theta=\pi/2$. The middle panel is taken at the scalar crossover scale, $T_{\rm pc}^{(S)}/\Lambda\simeq0.307$. This is distinct from the pseudoscalar scale in Fig.~\ref{fig:mass-phase}. Here $M_S=\bar\sigma+m/\sqrt2$ and
			$M_P=\bar\eta+m/\sqrt2$. The axes are not the total masses. The source-selected minimum moves continuously toward
			smaller auxiliary fields.}
		\label{fig:mass-contours}
	\end{figure*}
	
	\subsection{Mass phase and the condensates}
	\label{sec:mass-response}
	
	To study the response of a scalar-dominated condensate to the mass orientation, we now choose
	\begin{equation}
		\begin{gathered}
			G_S=1.50g_c,\qquad G_P=1.10g_c,\\
			C_4=0,\qquad m/\Lambda=0.08,
		\end{gathered}
		\label{eq:mass-benchmark}
	\end{equation}
	{at $\theta=0,\pi/2,\pi$. In contrast to the source-free example, the scalar coupling is now the larger one. Thus, this benchmark illustrates the response of a scalar-dominated system rather than the effect of adding a mass source to the previous benchmark with all other parameters fixed. The three phases that can emerge correspond to scalar, equally mixed, and pseudoscalar mass sources, respectively.}

	To distinguish the two channel responses, we track their magnitudes separately,
	\begin{equation}
		\begin{aligned}
			R_S(T)&=\frac{g_c}{\Lambda}|\langle\bar\psi\psi\rangle_T|,\\
			R_P(T)&=\frac{g_c}{\Lambda}|\langle\bar\psi i\gamma_5\psi\rangle_T|.
		\end{aligned}
		\label{eq:channel-magnitudes}
	\end{equation}
	For each smooth response, we then define a pseudocritical temperature by the maximum slope of either of the condensates defined in
	Eq.~(\ref{eq:channel-magnitudes}), i.e.,
	\begin{equation}
		T_{\rm pc}^{(i)}=\operatorname*{arg\,max}_T
		\left|\frac{\dd R_i(T)}{\dd T}\right|,
		\qquad i=S,P.
		\label{eq:Tpc}
	\end{equation}
	{\color{black}\textcolor{black}{These temperature scales mark smooth crossovers in the scalar and pseudoscalar condensates.}}
	This extends the usual scalar-condensate criterion~\cite{Fu:2019hdw,HotQCD:2019xnw,Borsanyi:2020fev,Wang:2024azs,Guan:2026rbs}. 
	
	
	\begin{figure*}[!t]
		\centering
		\threepanels{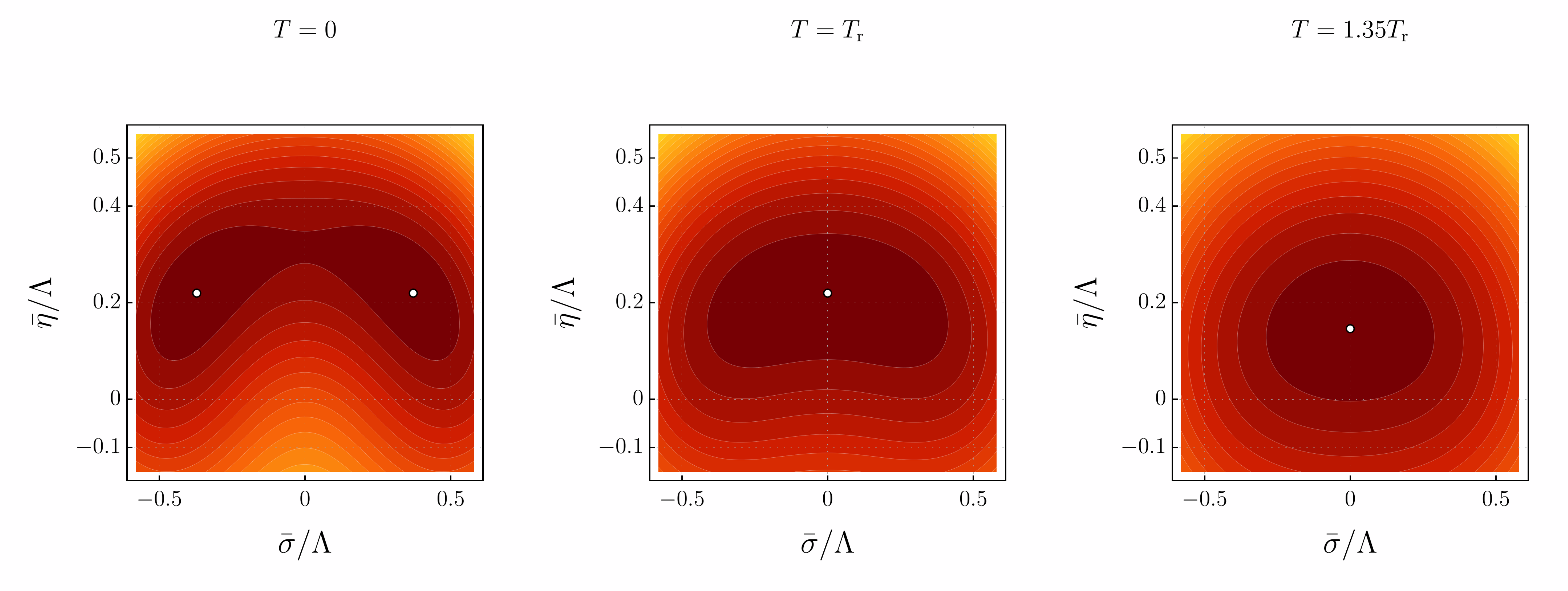}{fig:residual-contours}
		\caption{Potential contour levels in the plane $(\bar \sigma, \bar \eta)$  for $G_S=1.50g_c$, $G_P=1.10g_c$, $C_4=0$, $m/\Lambda=0.08$, $\theta=\pi$. Here $T_{\rm r}/\Lambda\simeq0.235$. At low temperature, two minima have opposite $\bar\sigma$ and the same nonzero $\bar\eta$. They merge in the scalar direction at $T_{\rm r}$, while the pseudoscalar response remains. The axes show auxiliary fields, with $M_S=\bar\sigma$ and $M_P=m+\bar\eta$.}
		\label{fig:residual-contours}
	\end{figure*}

	In Fig.~\ref{fig:mass-phase} we illustrate the condensates defined in Eq.~(\ref{eq:channel-magnitudes}). At $\theta=0$, only the scalar condensate forms, and its smooth decrease leaves parity unbroken. At $\theta=\pi/2$, both condensates decrease smoothly, although their largest slopes occur at different temperatures. By contrast, at $\theta=\pi$, the scalar condensate vanishes at a genuine reflection-symmetry transition, while the pseudoscalar condensate remains nonzero and reaches its largest slope immediately above that temperature. The corresponding scales are collected in Table~\ref{tab:thermal-scales}.
	
	In Fig.~\ref{fig:mass-contours} we show the potential contour levels in the plane $(\bar \sigma, \bar \eta)$ for the parameter benchmark values (cf. Eq.~\eqref{eq:mass-benchmark}) and for different temperature values. It illustrates  the smooth response for the case of $\theta=\pi/2$. Here, both mean fields are nonzero, and the mass source selects a single equilibrium minimum. As the temperature rises, this minimum moves continuously toward smaller mean field values. Thus, unlike the source-free case shown in Fig.~\ref{fig:CP-contours}, the crossover seem here does not involve the merger of two symmetry-related minima. Thus, the rapid decrease of the condensates describes a smooth thermal response, rather than the explicit restoration of ordinary $P/\CP$.
	
	
	At high temperature, the condensates approach zero, while the constituent masses $(M_S,M_P)$ approach $(m_S,m_P)$, as derived in Appendix~\ref{app:high-temperature-limit}. Consequently, the microscopic pseudoscalar source is not removed by the decrease of its condensate.
	

	\subsection{The residual transition at \texorpdfstring{$\theta=\pi$}{theta=pi}}
	\label{sec:residual-transition}
	
	A purely pseudoscalar mass source provides a useful exception to the smooth behavior seen for the condensates above. Although it explicitly breaks ordinary $P/\CP$, it leaves a reflection symmetry in the scalar direction. Then, the scalar condensate can vanish through a genuine transition. Indeed, with $m_S=C_4=0$ and $m_P=m$, the potential satisfies
	\begin{equation}
		\Omega_T(M_S,M_P)=\Omega_T(-M_S,M_P).
		\label{eq:residual-reflection}
	\end{equation}
	\begingroup
	{\color{black}This scalar reflection is distinct from ordinary $P/\CP$, which the pseudoscalar mass breaks explicitly. Its spontaneous breaking and restoration can be followed through the scalar condensate. We denote the restoration temperature by $T_{\rm r}$.}
	\endgroup
	
	{For $G_S>G_P$, the gap equations give a simple condition for this transition, as derived in Appendix~\ref{app:kernel}. Let $M_0(T)$ denote the source-free mass for the scalar coupling $G_S$, defined here by $\EuScript{K}_3(M_0(T),T)=1/G_S$. While the scalar condensate is nonzero, the same gap equations require
	\begin{equation}
		\begin{aligned}
			M_P&=\frac{G_Sm}{G_S-G_P},\\
			M_S&=\pm\sqrt{M_0^2(T)-M_P^2}.
		\end{aligned}
		\label{eq:residual-mass-components}
	\end{equation}
	The pseudoscalar condensate is therefore fixed at $\langle\bar\psi i\gamma_5\psi\rangle_T=-m/(G_S-G_P)$, explaining the plateau in the last panel of Fig.~\ref{fig:mass-phase}. The scalar condensate exists only while
	\begin{equation}
		M_0(T)>\left|\frac{G_Sm}{G_S-G_P}\right|.
		\label{eq:residual-existence}
	\end{equation}
	As $M_0(T)$ decreases with temperature, equality fixes $T_{\rm r}$, where the two scalar solutions meet. Thus, for $G_S>g_c$, a finite-temperature residual transition exists for sources in the range
	\begin{equation}
		|m|<\frac{G_S-G_P}{G_S}M_0(0).
		\label{eq:residual-source-range}
	\end{equation}
	For the benchmark in Eq.~\eqref{eq:mass-benchmark}, this gives $M_P/\Lambda=0.30$, a plateau at $R_P=0.20$, and $T_{\rm r}/\Lambda\simeq0.235$.}
	
	As a result, both condensates can be nonzero, as shown in Fig.~\ref{fig:residual-contours} for the same $\theta=\pi$ benchmark as the last panel of Fig.~\ref{fig:mass-phase}. At low temperature, the two minima have opposite scalar components but the same nonzero pseudoscalar component, $(\bar\sigma,\bar\eta)=(\pm\bar\sigma_0,\bar\eta_0)$. Hence, they are related by the scalar reflection, which changes $\bar\sigma$ while leaving $\bar\eta$ unchanged.
	
	As the temperature rises, the minima approach each other and merge at $\bar\sigma=0$ when $T=T_{\rm r}$. The scalar condensate then vanishes, as seen in Fig.~\ref{fig:mass-phase}. Unlike the source-free case in Fig.~\ref{fig:CP-contours}, the merged minimum is not at the origin. A pseudoscalar condensate remains because $m_P$ is nonzero. {\color{black}The transition restores the scalar reflection symmetry, while ordinary $P/\CP$ remains explicitly broken.}
	
	
	

	\subsection{Mixed interaction and pseudoscalar condensate}
	\label{sec:mixed-response}
	
	The mixed interaction induces both condensate components even when one mass source vanishes. To isolate its effect, we retain the mass benchmark $m/\Lambda=0.08$ and turn on the mixing coupling $C_4$, assuming for it the benchmark value
	\begin{equation}
		C_4=0.30g_c.
		\label{eq:mixed-benchmark}
	\end{equation}
	In particular, at $\theta=0$ the source $m_P$ vanishes, but the pseudoscalar gap equation becomes
	\begin{equation}
		\left[G_S-\Delta_G\EuScript{K}_3(M,T)\right]\bar\eta
		=C_4\bar\sigma.
		\label{eq:mixed-pseudoscalar-response}
	\end{equation}
	{Thus, through $C_4$, a nonzero scalar auxiliary field also induces a pseudoscalar condensate. Since the mixed interaction is already an explicit source, this response should not be identified with spontaneous $P/\CP$ breaking.}
	
	
	
	\begin{figure*}[!t]
		\centering
		\threepanels{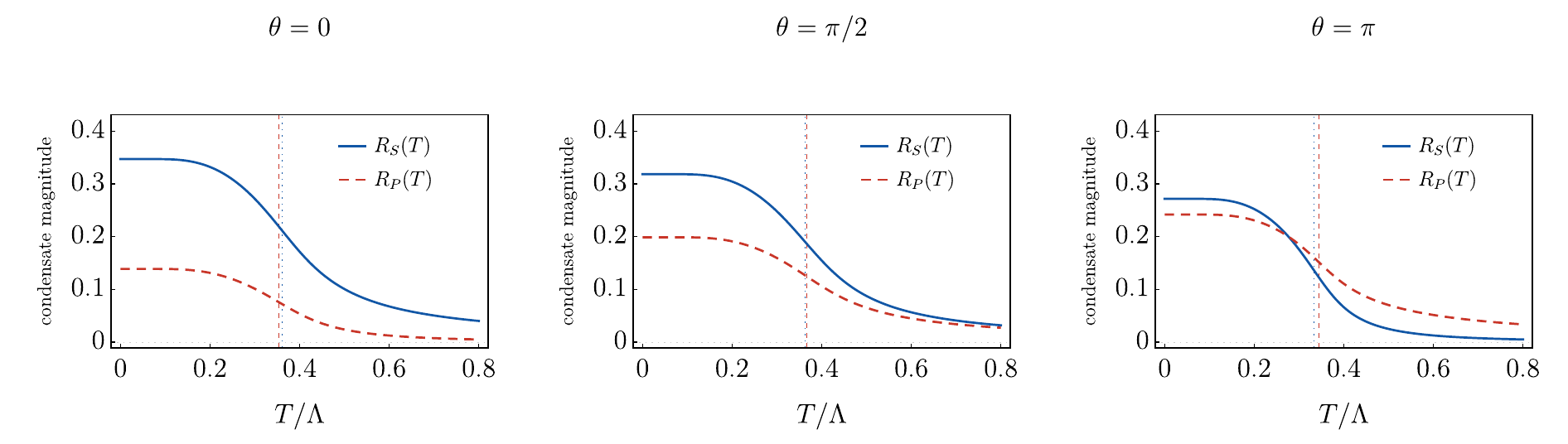}{fig:C4-mixing}
		\caption{Condensate response with scalar-pseudoscalar mixing for
			$G_S=1.50g_c$, $G_P=1.10g_c$, $C_4=0.30g_c$,
			$m/\Lambda=0.08$.
			Blue solid and red dashed curves show $R_S$ and $R_P$. Vertical lines in the same colors mark their respective maximum-slope temperatures. In the order $(T_{\rm pc}^{(S)}/\Lambda,T_{\rm pc}^{(P)}/\Lambda)$, these are approximately $(0.361,0.354)$, $(0.363,0.367)$, and $(0.334,0.345)$ for $\theta=0,\pi/2,\pi$, respectively, as listed in Table~\ref{tab:thermal-scales}. Both components are present even at $\theta=0$, where $m_P=0$. {\color{black}All displayed responses are smooth crossovers.}}
		\label{fig:C4-mixing}
	\end{figure*}
	
	Figure~\ref{fig:C4-mixing} shows both condensates decreasing smoothly for all three phases. In contrast to the $C_4=0$ endpoint, the mixed term breaks the scalar reflection in Eq.~\eqref{eq:residual-reflection}. The transition at $\theta=\pi$ is consequently replaced by a crossover. The phase dependence of the crossover scales reflects the changing alignment between the mass source and the preferred condensate direction.
	
\begin{figure*}[t]
		\centering
		\threepanels{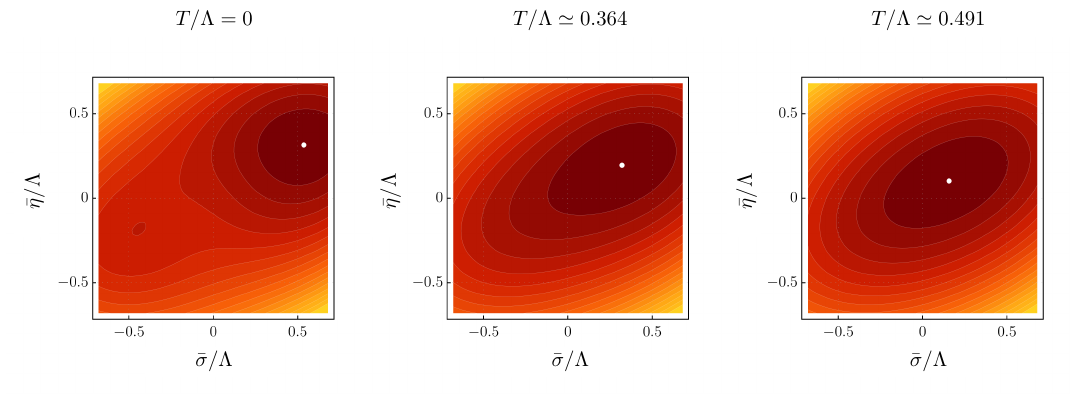}{fig:mixed-contours}
		\caption{Mixed-interaction potential contour levels in the plane $(\bar \sigma, \bar \eta)$  for $G_S=1.50g_c$, $G_P=1.10g_c$, $C_4=0.30g_c$, $m/\Lambda=0.08$, $\theta=\pi/2$. These panels illustrate the potential near and above the channel-resolved crossover temperatures, $T_{\rm pc}^{(S)}/\Lambda\simeq0.363$ and $T_{\rm pc}^{(P)}/\Lambda\simeq0.367$. Axes show auxiliary fields, with $M_S=m_S+\bar\sigma$ and $M_P=m_P+\bar\eta$. The tilted valley and moving minimum display mixing and thermal suppression.}
		\label{fig:mixed-contours}
	\end{figure*}

	{The potential contours in Fig.~\ref{fig:mixed-contours} show the same smooth behavior change in the two-field plane. Compared with the $C_4=0$ example in Fig.~\ref{fig:mass-contours}, the mixed interaction tilts the valley, and the temperature effect moves its single equilibrium minimum toward smaller auxiliary fields. There is no exchange of stability between separate minima. Thus, a weaker condensate response does not remove the explicit sources. In this way, mean-field theory describes the consequences of a prescribed mixed interaction, while the FRG analysis to be performed in the next section, will address how that interaction can be generated along the flow~\cite{Huang:2024ypj,Guan:2026tbd}.}
	
	Table~\ref{tab:thermal-scales} brings together the different channel-resolved scales. Only the source-free case and the residual-reflection endpoint correspond to genuine symmetry-restoring transitions. {\color{black}The other entries locate the fastest smooth response of each condensate. The momentum integrals and temperature scans were checked for numerical convergence, with changes in the crossover temperatures below $10^{-5}\Lambda$.}
	
	\begin{table*}[t]
		\color{black}
		\centering
		\renewcommand{\arraystretch}{1.10}
		\begin{tabular*}{0.80\textwidth}{@{\extracolsep{\fill}}lccl@{}}
			\toprule
			\multicolumn{4}{c}{(a) Symmetry-restoring transitions}\\
			\midrule
			Case & $\theta$ & Critical temperature & Restored symmetry\\
			\midrule
			Source-free $P$ channel & -- & $T_{\CP}/\Lambda=0.272$ & Ordinary $P$ and $\CP$\\
			Mass source & $\pi$ & $T_{\rm r}/\Lambda=0.235$ & \textcolor{black}{Scalar reflection}\\
			\bottomrule
		\end{tabular*}
		\par\medskip
		\begin{tabular*}{0.80\textwidth}{@{\extracolsep{\fill}}lccc@{}}
			\toprule
			\multicolumn{4}{c}{(b) Smooth crossovers}\\
			\midrule
			Case & $\theta$ & $T_{\rm pc}^{(S)}/\Lambda$ & $T_{\rm pc}^{(P)}/\Lambda$\\
			\midrule
			Mass source & $0$ & $0.326$ & --\\
			Mass source & $\pi/2$ & $0.307$ & $0.359$\\
			Mixed interaction & $0$ & $0.361$ & $0.354$\\
			Mixed interaction & $\pi/2$ & $0.363$ & $0.367$\\
			Mixed interaction & $\pi$ & $0.334$ & $0.345$\\
			\bottomrule
		\end{tabular*}
		\par\medskip
		\caption{\color{black}Mean-field temperatures for the benchmarks in Figs.~\ref{fig:CP-contours}, \ref{fig:mass-phase}, and \ref{fig:C4-mixing}. Panel (a) gives second-order transition temperatures and panel (b) gives crossover temperatures from the maximum slopes of the condensates. A dash denotes a vanishing condensate. At $\theta=\pi$ and $C_4=0$, the pseudoscalar response changes most rapidly just above $T_{\rm r}$, with no separate crossover.}
		\label{tab:thermal-scales}
	\end{table*}
	
	\section{Functional renormalization group at finite temperature}
	\label{sec:FRGsetup}
	
	{\setlength{\parskip}{0pt}We now focus on obtaining the finite-temperature FRG flow equations.\par}
	
	\vspace{-12pt}
	\subsection{General framework}
	\label{sec:frg-overview}
	
	In the preceding section, quark fluctuations were included through the one-loop fermion contribution to the mean-field effective potential associated with the model \eqref{eq:bare-action}. Here, we employ the FRG to investigate the RG-scale dependence of the model’s interactions at finite temperature. This approach incorporates fluctuation effects beyond a fixed one-loop calculation and offers a complementary view of how thermal fluctuations modify the consequences of explicit $\CP$-breaking sources. {Foundations and early developments of the FRG methodology are discussed in~\cite{Morris:1993qb,Reuter:1994average,Ellwanger:1993mw}. Reviews and lecture notes covering its methods, symmetries, and applications to gauge theories, statistical physics, and interacting fermions can be found in~\cite{Morris:1998elements,Berges:2000ew,Aoki:2000review,Bagnuls:2001review,Polonyi:2003lectures,Pawlowski:2007aspects,Gies:2012review,Delamotte:2012review,Sonoda:2007lectures,Igarashi:2010symmetry,Rosten:2012fundamentals,Braun:2011pp,Dupuis:2020fhh}.}
	
	The FRG construction implements the Wilsonian renormalization procedure, where the functional integral is not fully performed at once, but in a shellwise fashion, by integrating quantum fluctuations progressively according to their momentum scale. Starting from a theory defined at an UV scale $\Lambda$, fluctuations with momenta above a momentum scale $k$ are integrated out. The resulting scale dependence describes how the couplings evolve as the IR scale $k$ is lowered. In the limit $k\to0$, this is equivalent to the full functional integration of all fluctuations. Rather than excluding low-momentum modes with a sharp IR cutoff $k$, this procedure introduces a scale-dependent regulator term $\Delta S_k$ in the functional integral, which suppresses their contribution smoothly. As conventional in the FGR method, the regulator $\Delta S_k$ is chosen to be quadratic in the fields,
	\begin{equation}
		\Delta S_k[\Phi]=\frac{1}{2}\int \frac{\dd^4 p}{(2\pi)^4} \Phi^{\rm T}(-p)\cdot \mathcal R_k(p)\cdot \Phi(p)\,,
	\end{equation}
	where the matrix-valued regulator function $\mathcal R_k(p)$ suppresses fluctuations with momenta $|p| \lesssim k$, while leaving modes with $|p|>k$ essentially unaffected. {\color{black}For fixed momentum, the regulator vanishes as $k\to0$. The field content $\Phi$ is defined as}
	\begin{equation}
		\Phi(p)=\Big([\psi(p)]^{\rm T},\,
		\bar{\psi}(-p) \Big)^{\rm T}.
	\end{equation}
	The regulated generating functional is defined by
	\begin{equation}
		e^{W_k[\EuScript{J}]}
		=
		\int[\dd \Phi]\,
		e^{-S_{\mathrm E}[\Phi]-\Delta S_k[\Phi]	+\EuScript{J}\cdot\Phi}.
	\end{equation}
	Its Legendre transform, with the regulator term subtracted, defines
	the scale-dependent effective average action (or simply \textit{flowing action}) at supremum of the source,
	\begin{equation}
		\Gamma_k[\Phi]=\sup_\EuScript{J}\left(\EuScript{J}\cdot\Phi-W_k[\EuScript{J}]\right)-\Delta S_k[\Phi].
	\end{equation}
	Here, the source vector $\EuScript{J}$ is eliminated in favor of the corresponding average field $\Phi$. The regulator properties ensure that $\Gamma_k$ interpolates between the microscopic Euclidean action and the full effective action. With the UV initial condition imposed at $k=\Lambda$, one has $\Gamma_{k\to\Lambda}\cong S_{\rm E}$, while $\Gamma_{k \to 0}=\Gamma$. The latter is the one-particle-irreducible (1PI) effective action, whose functional derivatives generate the 1PI correlation vertices.

	The evolution of $\Gamma_k$ with the RG scale is governed by the Wetterich equation (or \textit{flow equation}) \cite{Wetterich:1992yh} (see also \cite{Ellwanger:1993mw,Morris:1993qb}),
	\begin{equation}
		\partial_t\Gamma_k[\Phi]=\frac12\,\mathrm{STr}\left[
		\left(\Gamma_k^{(2)}[\Phi]+ \mathcal R_k\right)^{-1}\cdot\partial_t \mathcal R_k\right],
		\label{eq:frg-wetterich}
	\end{equation}
	where $\partial_t=k\partial_k$ is the derivative with respect to the dimensionless RG ``time'' $t=\log(k/\Lambda)$. In this way, the UV limit $k\to\Lambda$ corresponds to $t\to0$, while the IR limit $k\to0$ corresponds to $t\to-\infty$. Thus, the flow toward the IR proceeds toward decreasing $t$.  The symbol STr denotes the functional supertrace, which comprises a trace over all discrete internal indices as well as an integration over continuous spacetime  variables, with appropriate minus signs accounting for fermionic statistics. In Eq.~(\ref{eq:frg-wetterich}), $\Gamma_k^{(2)}$ denotes the full field-dependent inverse propagator, i.e., the Hessian of $\Gamma_k$, defined as the second functional derivative with respect to the fields. Explicitly, this two-point function reads
	\begin{equation}
		\Gamma_{k,IJ}^{(2)}(p,q)=\frac{\overrightarrow{\delta}}{\delta \Phi_I^{\rm T}(-p)}\Gamma_k\frac{\overleftarrow{\delta}}{\delta \Phi_J(q)},
	\end{equation}
	where the indices $I$ and $J$ collectively label the color and spinor degrees of freedom of the fermionic fields. For an up to date review on the FRG and its modern applications, we refer the reader to~\cite{Dupuis:2020fhh}.
	
	A few features of Eq.~\eqref{eq:frg-wetterich} are worth emphasizing: the operator $\left(\Gamma_k^{(2)}[\Phi]+ \mathcal R_k\right)^{-1}$ is the field-dependent, regulated propagator, \ie, it is dressed with quantum fluctuations above $k^2$. Also, the flow equation is exact: no approximation is made in its derivation. Moreover, once we integrate \eqref{eq:frg-wetterich} down to $k= 0$, we are led to $\Gamma$. In this sense, solving the flow equation \eqref{eq:frg-wetterich} and taking the limit $k\rightarrow 0$ is formally equivalent to performing the functional integral exactly.
	
	In practice, Eq.~\eqref{eq:frg-wetterich} is a functional differential equation that cannot generally be solved in full. We can then specify
	an ansatz for $\Gamma_k$ that contains a selected set of operators and
	project the flow \eqref{eq:frg-wetterich} onto their scale-dependent coefficients. This approximation, known as a \textit{truncation}, neglects the feedback of operators outside the chosen ansatz.
	
	\subsection{Truncated effective action}
	\label{sec:frg-truncation}
	
	We approximate $\Gamma_k$ by retaining the operator structure of the Euclidean action in Eq.~\eqref{eq:bare-action}, while allowing its couplings to run with the RG scale $k$. Explicitly,
	\begin{align}
		\Gamma_k[\Phi]\simeq&\int_x\Big\{Z_{\psi}\bar\psi\big(
		\gamma_\mu\partial_\mu+m e^{i\gamma_5 \theta/2}\big)\psi-\frac{G_{S}}2(\bar\psi\psi)^2
		\nonumber\\[-1mm]
		&-\frac{G_{P}}2(\bar\psi i\gamma_5\psi)^2-C_{4}(\bar\psi\psi)(\bar\psi i\gamma_5\psi)\Big\},
		\label{eq:frg-ansatz}
	\end{align}
	Throughout this section, the set of parameters $m, \theta, G_S, G_P, C_4$ denote scale-dependent quantities that evolve with the RG scale $k$. Here, we work at leading order in the derivative expansion. We keep the fermion kinetic term in its standard form, setting $Z_{\psi}=1$, and evaluate the interaction vertices at vanishing external momenta. This corresponds to the local potential approximation (LPA).
	
	{\color{black}
		Note that Eq.~\eqref{eq:frg-ansatz} retains the scalar, pseudoscalar, and mixed interactions to follow the condensate dynamics and $P/\CP$ mixing in the same sector as the mean-field analysis. A Fierz-complete treatment would also account for independent vector $(\bar\psi\gamma_\mu\psi)^2$, axial-vector $(\bar\psi\gamma_\mu\gamma_5\psi)^2$, and tensor $(\bar\psi\sigma_{\mu\nu}\psi)^2$ interactions, whose four-fermion couplings have the same canonical dimension as those retained here~\cite{Braun:2017srn,Braun:2018fierz2,Braun:2020fierz3}. At finite temperature, their temporal and spatial components may also require separate couplings. Such a treatment would therefore be more comprehensive. Nevertheless, studies of enlarged truncations show that additional directions can remain IR-irrelevant near interacting fixed points~\cite{Braun:2017srn,Aoki:1999critical,Guan:2026tbd}\footnote{\color{black}A relevant (irrelevant) direction describes a small departure from a fixed point that grows (decays) toward the IR. These directions can involve combinations of couplings, rather than a single interaction channel. Their classification is explained in Sec.~\ref{sec:zero-T-case}.}. This motivates focusing on the coupled scalar-pseudoscalar sector to isolate its thermal behavior, without assuming that the omitted channels have no quantitative effect on the phase boundary.}
	

	It is also useful to comment on the four-fermion interactions as far as phase transitions are concerned.
	A divergence of the pointlike four-fermion couplings signals
	an instability toward condensate formation in the coupled
	scalar-pseudoscalar sector. When the relevant chiral symmetry
	is exact, this provides a diagnostic of the onset of spontaneous
	chiral symmetry breaking~\cite{Braun:2011pp}. In the presence of
	explicit breaking sources, it instead identifies a strong-coupling	scale associated with condensate dynamics. \textcolor{black}{Since the present fermionic flow does not determine the effective	potential for the condensates, the divergence alone does not identify the equilibrium state beyond the instability. To study that regime, the flow can be extended using dynamical hadronization~\cite{Gies:2002boundstates,Floerchinger:2009composites,Braun:2016dynamicalQCD}	or a weak-solution formulation of the
		FRG~\cite{Guan:2026weak}.}
	
	\subsection{Flow equations}
	\label{sec:frg-flows}
	
	We now proceed to apply the Wetterich equation (\ref{eq:frg-wetterich}) to the effective action \eqref{eq:frg-ansatz} to derive the RG flow of the couplings, \ie, their corresponding beta functions. For an explicit evaluation of the flow equation, a specific choice of the regulator matrix is required. At finite temperature, the supertrace appearing in the flow equation includes a sum over fermionic Matsubara frequencies. This sum can be evaluated analytically provided that the frequency dependence of the propagators and vertices is retained at its classical form. The explicit derivation of the corresponding $\beta$-functions is provided in Appendix~\ref{app:frg-derivation}. We then employ a regulator that acts only on the spatial momentum components, such that $\mathcal{R}_k(p)\longrightarrow \mathcal{R}_k(\vb*{p}),$
	while leaving the temporal momentum component unregularized. More explicitly, we choose the regulator matrix
	\begin{equation}
		\mathcal{R}_k(\vb*{p})=
		\begin{pmatrix}
			0 &
			i\slashed{\vb*{p}}^{\mathrm T}
			r_k^\psi(\vb*{p}^{\,2}/k^2)
			\\[1ex]
			i\,\slashed{\vb*{p}}\,
			r_k^\psi(\vb*{p}^{\,2}/k^2)
			&
			0
		\end{pmatrix}\delta_{p,q},
	\end{equation}
	where we employ the Litim-type shape function \cite{Litim:2001up,Litim:2006ag}
	\begin{equation}
		r_k^{\psi}(\vb*{p}^{\,2}/k^2)=\left(\sqrt{\frac{k^2}{\vb*{p}^2}}-1\right)\Theta(k^2-\vb*{p}^2),
		\label{eq:frg-regulator}
	\end{equation}
	and define $\delta_{p,q}=\beta(2\pi)^3\delta_{n_p,n_q}\delta^{(3)}(\vb*{p}-\vb*{q})$.
	
	We retain the leading color-enhanced fermionic contributions in the
	large-$N_c$ expansion and set $N_c=3$ for the numerical benchmarks.
	To investigate the fixed-point structure and RG trajectories, we
	introduce the dimensionless variables
	\begin{align}
		\tilde G_S&=k^2{G_S},\qquad
		\tilde G_P=k^2{G_P},\qquad
		\tilde C_4=k^2{C_4},\nonumber\\*
		\tilde m&=\frac{{m}}{k},\qquad
		\tau=\frac{T}{k}.
		\label{eq:frg-dimensionless}
	\end{align}
	The phase $\theta$ is dimensionless. At finite temperature, we introduce
	$\tau=T/k$ as an additional running variable, with the physical
	temperature $T$ held fixed along each RG trajectory. Its scale
	dependence is therefore purely canonical,
	satisfying $\partial_t\tau=-\tau$.
	
	At leading order in the large-$N_c$ expansion, the complete set of coupled flow equations for these dimensionless couplings are
	\begin{widetext}
		\begin{subequations}\label{eq:frg-beta-functions}
			\begin{align}
				\partial_t\tilde G_S=&2\tilde G_S-8N_c\Big[
				(\tilde G_S^2+\tilde C_4^2)\widehat{\mathcal A}
				-2\tilde m^2(\tilde G_S^2-\tilde C_4^2)
				\cos\theta\,\widehat{\mathcal B}
				-4\tilde m^2\tilde G_S\tilde C_4
				\sin\theta\,\widehat{\mathcal B}\Big],
				\\
				\partial_t\tilde G_P=&2\tilde G_P-8N_c\Big[
				(\tilde G_P^2+\tilde C_4^2)\widehat{\mathcal A}
				+2\tilde m^2(\tilde G_P^2-\tilde C_4^2)
				\cos\theta\,\widehat{\mathcal B}
				-4\tilde m^2\tilde G_P\tilde C_4
				\sin\theta\,\widehat{\mathcal B}\Big],
				\\
				\partial_t\tilde C_4=&2\tilde C_4-8N_c\Big[
				(\tilde G_S+\tilde G_P)\tilde C_4\widehat{\mathcal A}
				-2\tilde m^2(\tilde G_S-\tilde G_P)
				\tilde C_4\cos\theta\,\widehat{\mathcal B}
				-2\tilde m^2(\tilde G_S\tilde G_P+\tilde C_4^2)
				\sin\theta\,\widehat{\mathcal B}\Big],
				\\
				\partial_t\tilde m=&-\tilde m
				-4N_c\tilde m\Big[
				(\tilde G_S+\tilde G_P)
				+(\tilde G_S-\tilde G_P)\cos\theta
				+2\tilde C_4\sin\theta\Big]\widehat{\mathcal M},
				\\
				\partial_t\theta=&8N_c\Big[
				(\tilde G_S-\tilde G_P)\sin\theta
				-2\tilde C_4\cos\theta\Big]\widehat{\mathcal M}.
			\end{align}
		\end{subequations}
	\end{widetext}
	In the above equations, the explicit expressions for the dimensionless threshold functions
	$\widehat{\mathcal A}(\tau,\tilde m)$, $\widehat{\mathcal B}(\tau,\tilde m)$ and $\widehat{\mathcal M}(\tau,\tilde m)$ are given in Appendix~\ref{app:frg-derivation}.
	
	We remark that the flow preserves the $\CP$-even subspace $\tilde C_4=\theta=0$, and $\partial_t \theta=0$ whenever
	$\tilde G_S=\tilde G_P$ and $\tilde C_4=0$.
	Furthermore, a vanishing UV coupling need not remain zero, as can be seen from the beta function of $\tilde G_P$:
	\begin{equation}
		\left.\partial_t\tilde G_P\right|_{\tilde m=\tilde G_P=0}
		=-8N_c\tilde C_4^2\widehat{\mathcal A}(\tau,0).
	\end{equation}
	Thus, a nonzero mixed interaction $\tilde C_4$ generates a pseudoscalar coupling along the RG flow, even if $\tilde G_P$ vanishes at the UV scale. Therefore, we can retain the complete coupled system for the principal
	results and use reduced flows only as auxiliary diagnostics.
	
	
	\section{FRG Results}
	\label{sec:frg-results}
	
	{Let us now discuss the thermal flows and their connection with the mean-field results.}
	
	\subsection{Fixed-point structure}
	
	{\color{black}We first study the fixed-point structure of an auxiliary two-channel projection of Eq.~\eqref{eq:frg-beta-functions} in the $(\tilde G_S,\tilde C_4)$ plane, rather than classify the complete coupled system. This projection isolates how the scalar and mixed interactions shape the boundary between regular flows and finite-scale instabilities. We work in the chiral limit, $\tilde m=0$. Since the mass flow is proportional to $\tilde m$ itself, an initially vanishing mass remains zero within the pointlike approximation~\cite{tHooft:1979rat,Wells:2013tta}. This approximation ceases to be valid when the four-fermion couplings diverge. We begin with the vacuum fixed points at $T=0$ and then examine how temperature modifies the flow.}
	
	\subsubsection{Zero temperature case}\label{sec:zero-T-case}
	
	{\color{black}For this projection, we set $\tilde G_P=\theta=0$ and drop their flow equations.} The massless interaction flow equations are independent of $\theta$, so this choice does not affect the retained flows. In the massless vacuum,
	$\widehat{\mathcal A}(0,0)=1/(24\pi^2)$, and the retained flows become
	\begin{subequations}\label{eq:frg-two-channel-vacuum}
		\begin{align}
			\partial_t\tilde G_S
			&=2\tilde G_S-\frac{2}{G_*}
			(\tilde G_S^2+\tilde C_4^2),\\
			\partial_t\tilde C_4
			&=2\tilde C_4\left(1-\frac{\tilde G_S}{G_*}\right),
			\qquad G_*\equiv\frac{6\pi^2}{N_c}.
		\end{align}
	\end{subequations}
	The fixed points are determined by the simultaneous conditions
	$\partial_t\tilde G_S=\partial_t\tilde C_4=0$.
	The mixed-channel condition implies either $\tilde C_4^*=0$
	or $\tilde G_S^*=G_*$. Substituting the latter into the
	scalar-channel condition again requires $\tilde C_4^*=0$.
	Thus, this partial system admits exactly two real fixed points.
	In the coordinate ordering
	$\mathbf{FP}^{(0)}_i=(\tilde G_S^*,\tilde C_4^*)$, these are
	the Gaussian and non-Gaussian fixed points,
	\begin{equation}
		\mathbf{FP}^{(0)}_1=(0,0),\qquad
		\mathbf{FP}^{(0)}_2=(G_*,0),
		\label{eq:frg-two-channel-fixed-points}
	\end{equation}
	respectively.
	
	\begin{figure*}[!t]
		\begin{minipage}[t]{0.49\textwidth}
			\centering
			\includegraphics[width=\linewidth]{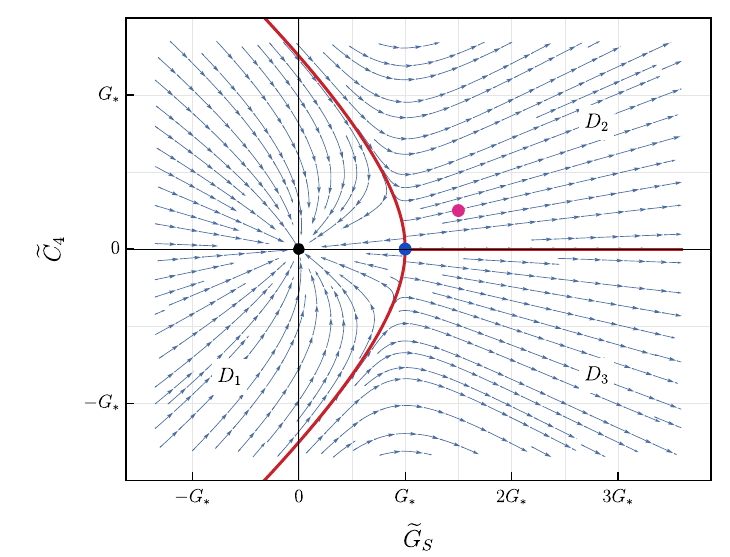}
			\caption{Vacuum two-channel flow for $\tilde m=0$, with $\tilde G_P$ and its beta function dropped. The pink point denotes the UV initial condition $(\tilde G_S,\tilde C_4)=(3G_*/2,G_*/4)$, where $G_*=2\pi^2$. Black and blue points mark the Gaussian and interacting fixed
				points, respectively. Thin blue arrows indicate the IR direction, whereas red solid curves are separatrices. The domains $D_1,D_2,D_3$ are defined
				in the text.}
			\label{fig:frg-two-channel-vacuum}
		\end{minipage}\hfill
		\begin{minipage}[t]{0.49\textwidth}
			\centering
			\includegraphics[width=\linewidth]{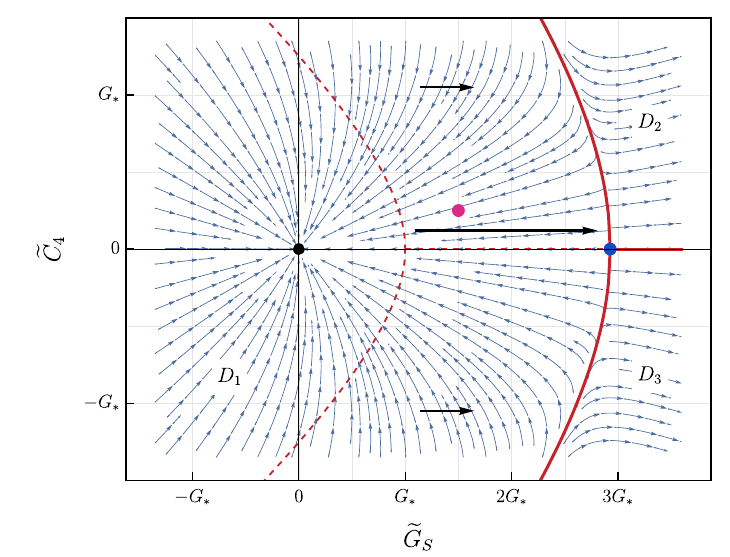}
			\caption{\textcolor{black}{Two-channel portrait at frozen $\tau=0.5$, with $\tilde m=0$, and the same projection and UV couplings as in Fig.~\ref{fig:frg-two-channel-vacuum}. Black and blue dots mark the Gaussian and non-Gaussian pseudo-fixed points, and pink marks the UV point. Blue arrows indicate IR evolution. Solid and dashed red curves show the thermal and vacuum separatrices, respectively. Black arrows indicate the thermal shift.}}
			\label{fig:frg-two-channel-thermal}
		\end{minipage}
	\end{figure*}
	
	The local behavior near the fixed point $g_i^*=\{\tilde G_S^*,\tilde C_4^*\}$ follows from linearizing
	the flow equations $\partial_t g_i=\beta_i(g)$.
	Writing $\delta g_i=g_i-g_{i}^*$, we obtain
	\begin{equation}
		\partial_t\delta g_i
		=\sum_j B_{ij}\delta g_j+O(\delta g^2),	
	\end{equation}
	where $B_{ij}=\partial_{g_j}\beta_i\big|_{g=g_*}$ is the stability matrix. The linearized flow has the solution
	\begin{equation}
		g_i(t)=g_{i}^*
		+\sum_a c_a V_i^{(a)}e^{-\vartheta_a t},
		\label{eq:frg-linearized-solution}
	\end{equation}
	where $c_a$ are integration constants and $V^{(a)}$ is the $a$-th eigenvector of $B_{ij}$, satisfying $B_{ij}V_j^{(a)}=-\vartheta_a V_i^{(a)}$. When the critical exponents $\vartheta_a>0$, as $t$ decreases toward the IR,
	the flow moves away from the fixed point along that
	eigendirection, provided that the corresponding
	perturbation amplitude $c_a$ is nonzero. Conversely, as $t$ increases, the flow approaches the fixed point. Such a direction is locally UV-attractive and IR-repulsive. Therefore, positive critical exponents characterize the so-called \textit{relevant directions}.
	
	For $\vartheta_a<0$, the behavior is reversed. The distance between $g_i$ and $g_i^*$ decreases toward the IR, so the flow approaches the fixed point along that direction. These directions are called \textit{irrelevant} and are locally UV-repulsive and IR-attractive. A vanishing exponent defines a marginal direction, whose behavior requires a nonlinear analysis. In the present two-channel system, $B_{ij}$ is exactly diagonal at both fixed points because $\tilde C_4^*=0$. We may then choose $V_i^{(a)}=\delta_{ia}$ and obtain the critical exponents directly from the diagonal entries, $\vartheta_i=-B_{ii}$, without summation over $i$.
	
	At the Gaussian fixed point, the critical exponents are $(-2,-2)$,
	so both directions are IR-attractive. At the non-Gaussian
	fixed point, they are $(2,0)$, identifying an IR-repulsive
	scalar direction and a linearly marginal mixed direction.
	
	For illustration, in Fig.~\ref{fig:frg-two-channel-vacuum} we show the vacuum flow in the $(\tilde G_S,\tilde C_4)$ plane. The black and blue points mark the Gaussian and non-Gaussian fixed points, respectively. The pink point denotes the UV initial condition
	$\tilde G_S(0)=3G_*/2$ and $\tilde C_4(0)=G_*/4$, with $G_*=2\pi^2$, and the arrows indicate evolution toward the IR.
	The red critical branches form the separatrix between initial conditions flowing toward the Gaussian point in $D_1$ and those developing a finite-scale divergence. Trajectories on these branches approach the non-Gaussian point toward the IR, with the attraction determined by nonlinear terms. Departures from this boundary
	instead grow toward the IR. The domains $D_2$ and $D_3$ are separated by the scalar-axis trajectory $\tilde C_4=0$, $\tilde G_S>G_*$.
	The RG flow preserves $\tilde C_4=0$, so a trajectory starting on this axis remains there. {\color{black}These domains distinguish divergent flows with opposite signs of the mixed coupling. They describe flow behavior and do not identify distinct thermodynamic phases.}
	
	The selected UV point lies in $D_2$, and its trajectory develops a pole at a finite scale $\textcolor{black}{k_{\rm pole}}=\Lambda e^{\textcolor{black}{t_{\rm pole}}}$. For our benchmark, this gives $\textcolor{black}{t_{\rm pole}}\simeq-\textcolor{black}{0.524}$, corresponding to $\textcolor{black}{k_{\rm pole}}/\Lambda\simeq\textcolor{black}{0.592}$. The scalar coupling already exceeds $G_*$, while the $\tilde C_4^2$ contribution reinforces its growth toward the IR. Thus, the portrait illustrates
	the role of the mixed interaction in the vacuum instability and identifies the boundary between regular and divergent flows.
	
	\begin{figure*}[!t]
		\centering
		\begin{minipage}[t]{0.49\textwidth}
			\centering
			\includegraphics[width=\linewidth]{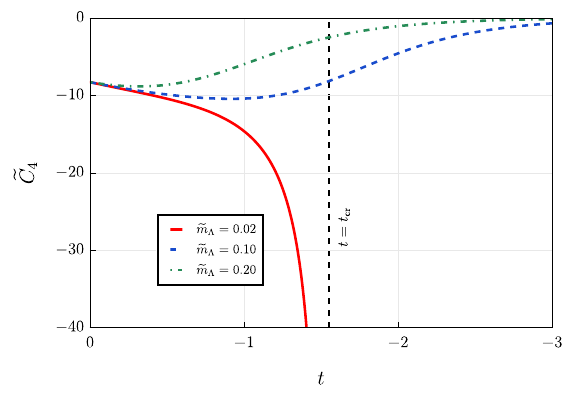}
			\par\vspace{2pt}{\small(a)}
		\end{minipage}\hfill
		\begin{minipage}[t]{0.49\textwidth}
			\centering
			\includegraphics[width=\linewidth]{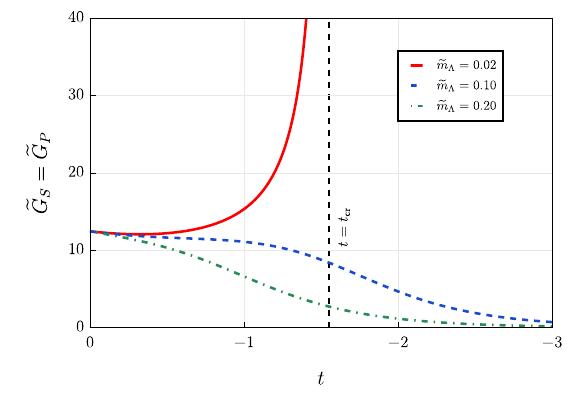}
			\par\vspace{2pt}{\small(b)}
		\end{minipage}
		\caption{Vacuum mass dependence of (a) $\tilde C_4$ and (b) $\tilde G_S=\tilde G_P$, with $\tilde G_S(0)=\tilde G_P(0)=0.63G_*$, $\tilde C_4(0)=-0.42G_*$, $G_*=2\pi^2$, and $\theta(0)=\pi/2$. In both panels, the flow proceeds toward the IR from left to right. Dashed vertical lines mark the common pole for the smallest mass, beyond which that trajectory is not continued. The two larger-mass trajectories exhibit no pole over the numerically explored interval $-18\leq t\leq0$, well beyond the range shown.}
		\label{fig:frg-vacuum-mass-benchmark}
	\end{figure*}
	
	\subsubsection{Finite temperature case}
	
	{\color{black}We examine thermal effects in two complementary ways. Portraits at fixed $\tau=T/k$ show how temperature shifts the fixed points and the boundaries between different flows. Integrations at fixed physical $T$ instead determine the temperature above which finite-scale poles disappear for the chosen UV couplings. We first construct the frozen-$\tau$ portraits. The Gaussian fixed point remains at the origin, while the interacting point moves with $\tau$. We then refer to the latter as a non-Gaussian pseudo-fixed point, with coordinates}
	\begin{equation}
		(\tilde G_S^*(\tau),\tilde C_4^*(\tau))
		=\left(\frac{1}{4N_c\widehat{\mathcal A}(\tau,0)},0\right),
		\label{eq:frg-two-channel-thermal-zero}
	\end{equation}
	with
	\begin{equation}
		\frac{1}{4N_c\widehat{\mathcal A}(\tau,0)}
		={\frac{G_*}{\displaystyle
				\tanh\!\left(\frac{1}{2\tau}\right)
				-\frac{1}{2\tau}\operatorname{sech}^2\!\left(\frac{1}{2\tau}\right)}}.
		\label{eq:frg-two-channel-thermal-position}
	\end{equation}
	Here $\tau>0$, and the vacuum value $G_*$ is recovered as $\tau\to0$. At fixed $\tau$, the retained equations have the same form as Eq.~\eqref{eq:frg-two-channel-vacuum}, with $G_*$ replaced by $\tilde G_S^*(\tau)$. Their local stability classification is consequently unchanged.
	
	Figure~\ref{fig:frg-two-channel-thermal} shows the portrait at $\tau=0.5$, for which $\tilde G_S^*(0.5)\simeq\textcolor{black}{2.93}G_*$. The pink point retains the vacuum UV initial values $\tilde G_S(0)=3G_*/2$ and $\tilde C_4(0)=G_*/4$. Thermal suppression of the fermionic threshold function shifts the non-Gaussian pseudo-fixed point and the critical branches toward larger scalar coupling. The unchanged initial point now lies in $D_1$, and its frozen-$\tau$ trajectory approaches the Gaussian fixed point rather than developing the vacuum pole. {\color{black}The black arrows show how the pseudo-fixed point and critical branches shift as $\tau$ increases, not the direction of RG evolution. The dashed red curves reproduce the vacuum separatrices for comparison. Together, these changes show how thermal effects weaken the instability.}
	
	\begin{figure*}[!t]
		\centering
		\begin{minipage}[t]{0.49\textwidth}
			\centering
			\includegraphics[width=\linewidth]{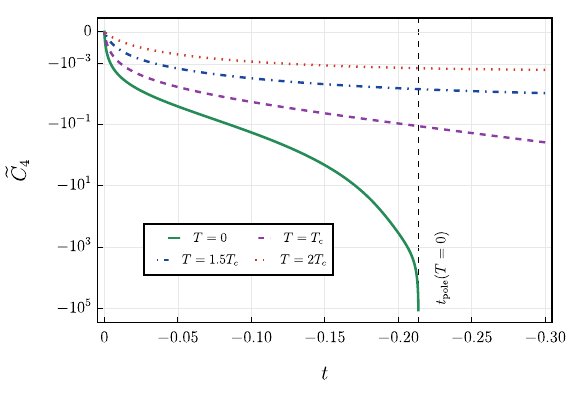}
			\par\vspace{2pt}{\small(a)}
		\end{minipage}\hfill
		\begin{minipage}[t]{0.49\textwidth}
			\centering
			\includegraphics[width=\linewidth]{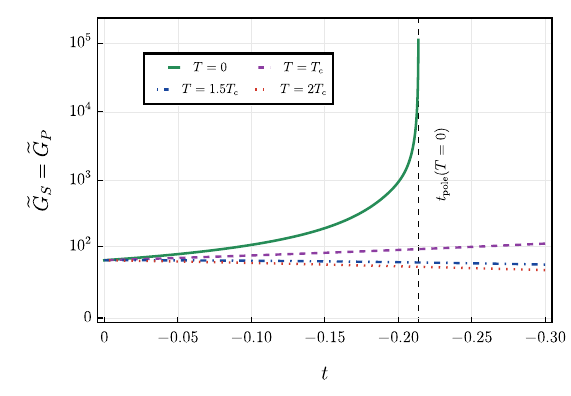}
			\par\vspace{2pt}{\small(b)}
		\end{minipage}
		\caption{Thermal evolution of (a) the generated mixed coupling $\tilde C_4$ and (b) $\tilde G_S=\tilde G_P$. The complete coupled equations are integrated with $\tilde G_S(0)=\tilde G_P(0)=3G_*$, $\tilde C_4(0)=0$, $\tilde m(0)=0.02$, $\theta(0)=\pi/2$, and $G_*=2\pi^2$. Here $\textcolor{black}{T_c}$ marks the disappearance of finite-scale poles, with $\textcolor{black}{T_c}/\Lambda\simeq0.338$. Each trajectory has fixed physical $T$, while $\tau=T/k$ runs. Dashed vertical lines mark the vacuum pole at $\textcolor{black}{t_{\rm pole}}\simeq-\textcolor{black}{0.214}$, beyond which that trajectory is not continued. The trajectories at $1.5\textcolor{black}{T_c}$ and $2\textcolor{black}{T_c}$ were followed down to $t=-18$, with no pole detected beyond the displayed range.}
		\label{fig:frg-generated-thermal-flows}
	\end{figure*}
	
	{\color{black}At fixed physical $T$, however, $\tau=T/k$ increases as $k$ decreases, according to $\partial_t\tau=-\tau$. The pseudo-fixed point and the critical curves therefore move during the flow. A red separatrix separates regular and divergent trajectories only when $\tau$ is held fixed. A trajectory starting on that curve need not remain on it when evolved at fixed $T$.}
	
	{\color{black}For the fixed-$T$ integrations, we use the UV couplings of Fig.~\ref{fig:frg-two-channel-vacuum} and obtain $\textcolor{black}{T_c^{\textmd{two-channel}}}/\Lambda\simeq\textcolor{black}{0.201}$, separating trajectories that develop a finite-scale pole from those approaching the Gaussian fixed point.} As $T$ approaches $\textcolor{black}{T_c^{\textmd{two-channel}}}$ from below, the instability scale is driven towards zero.
	Above this boundary, thermal suppression prevents the strong-coupling growth of both retained interactions, whose dimensionless couplings vanish in the infrared. This behavior realizes, within the present two-channel approximation, the competition between interaction-driven instability and thermal decoupling discussed in \cite{Braun:2018fierz2}. In particular, a nonzero mixed UV interaction does not enforce a finite-scale instability
	at arbitrarily high temperature. Instead, the resulting boundary characterizes the thermal suppression of the coupled scalar-pseudoscalar instability.
	
	When $\tilde C_4(0)=0$, the fixed-point and separatrix
	structure is unchanged, while the selected trajectory remains
	on the scalar axis because the projected flow preserves
	$\tilde C_4=0$. Then, with the same scalar UV coupling, the
	vacuum divergence and its suppression at sufficiently high
	temperature persist, although the corresponding instability
	scale and \textcolor{black}{critical temperature} change.
	
	\begin{figure}[!t]
		\centering
		\includegraphics[width=\columnwidth]{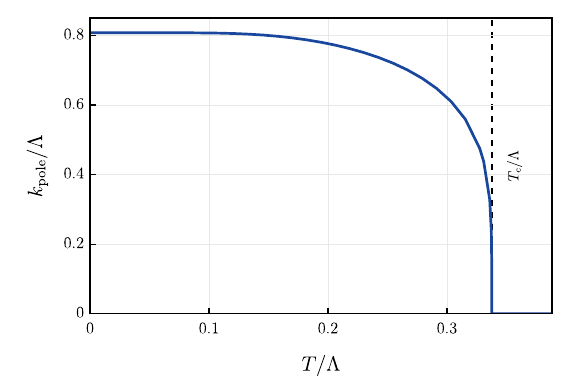}
		\caption{Temperature dependence of the pole scale for the complete coupled flow, with $\tilde G_S(0)=\tilde G_P(0)=3G_*$, $\tilde C_4(0)=0$, $\tilde m(0)=0.02$, $\theta(0)=\pi/2$, and $G_*=2\pi^2$, as in Fig.~\ref{fig:frg-generated-thermal-flows}. Each flow is integrated at fixed physical temperature. The dashed line marks $\textcolor{black}{T_c}/\Lambda\simeq\textcolor{black}{0.338}$. {\color{black}The zero values above $\textcolor{black}{T_c}$ indicate that no finite-scale pole is found.}}
		\label{fig:frg-thermal-pole-scale}
	\end{figure}
	
	\begin{figure*}[!t]
		\centering
		\begin{minipage}[t]{0.49\textwidth}
			\centering
			\includegraphics[width=\linewidth]{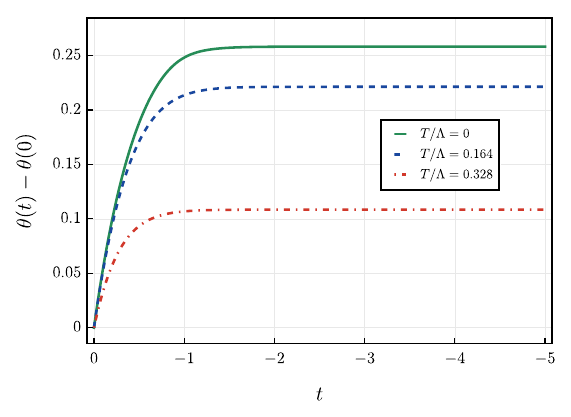}
			\par\vspace{2pt}{\small(a) $\theta(0)=0$}
		\end{minipage}\hfill
		\begin{minipage}[t]{0.49\textwidth}
			\centering
			\includegraphics[width=\linewidth]{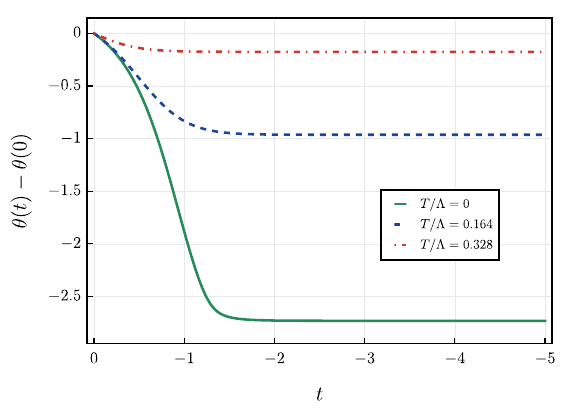}
			\par\vspace{2pt}{\small(b) $\theta(0)=\pi$}
		\end{minipage}
		\caption{Thermal evolution of the mass-phase shift, in radians, for initial phases (a) $\theta(0)=0$ and (b) $\theta(0)=\pi$. Both panels use $G_*=2\pi^2$, $\tilde G_S(0)=1.05G_*$, $\tilde G_P(0)=0$, $\tilde C_4(0)=0.15G_*$, and $\tilde m(0)=0.02$. The physical temperatures are fixed along each trajectory. All five equations are integrated, with no pole detected down to $t=-8$, beyond the displayed interval. The vertical ranges differ to resolve the different phase shifts.}
		\label{fig:frg-phase-evolution}
	\end{figure*}
	
	\subsection{RG flow of the four-fermion couplings}
	
	We next examine how an explicit mass affects the vacuum instability in the complete coupled system. 
	
	In Fig.~\ref{fig:frg-vacuum-mass-benchmark}, we compare three UV masses at fixed UV interactions, using the benchmark specified in the caption. All five flow equations are retained, and $\tilde G_S=\tilde G_P$ and $\theta=\pi/2$ are preserved by the coupled flow. The trajectory with $\tilde m(0)=0.02$ develops a common pole at $\textcolor{black}{t_{\rm pole}}\simeq-\textcolor{black}{1.55}$, whereas the two larger masses yield regular flows. Thus, a finite UV mass does not exclude an interaction-driven instability, but stronger mass-induced decoupling can prevent its formation.
	
	We now keep the UV mass fixed and vary the physical temperature, using a different benchmark with stronger diagonal interactions and $\tilde C_4(0)=0$, as specified in Fig.~\ref{fig:frg-generated-thermal-flows}. Although the mixed coupling initially vanishes, the mass-dependent term in its flow equation generates it. Its initial beta function is positive, so evolution toward decreasing $t$ produces a negative $\tilde C_4$. The scalar and pseudoscalar couplings remain equal and grow together, reaching the same vacuum pole as the mixed interaction.
	
	{\color{black}For these UV conditions, we denote by $\textcolor{black}{T_c}$ the temperature separating flows that develop a finite-scale pole from those that remain finite toward the infrared. This scale is determined from the RG flow, rather than from the thermodynamic potential. Unlike $\textcolor{black}{T_c^{\textmd{two-channel}}}$ above, $T_c$ refers to the complete coupled system and a different UV benchmark.} {We describe how the pole scale and $T_c$ are determined numerically in Appendix~\ref{app:frg-pole-numerics}.}
	
	As the temperature increases, fermion fluctuations contribute less to the flow, slowing the growth of the couplings. For these UV conditions, the finite-scale divergence disappears above $\textcolor{black}{T_c}/\Lambda\simeq\textcolor{black}{0.338}$. The trajectories at $1.5\textcolor{black}{T_c}$ and $2\textcolor{black}{T_c}$ remain finite, and their dimensionless couplings approach zero when followed further into the IR. The curve at $T=\textcolor{black}{T_c}$ separates this behavior from flows that develop a finite-scale pole.
	{\color{black}Thus, a nonzero $\tilde C_4$ is generated dynamically along the flow, even though its UV value vanishes.} A sufficiently high temperature prevents its growth into a finite-scale divergence.
	For the regular trajectories at $1.5\textcolor{black}{T_c}$ and $2\textcolor{black}{T_c}$, the decrease of the dimensionless couplings at low scales mainly reflects their canonical scaling. Once thermal suppression makes fermion fluctuations contribute little to the flow, the dimensionful couplings nearly stop running, while their dimensionless counterparts continue to decrease through the factor $k^2$. In particular, $\tilde C_4=k^2C_{4}$ approaches zero even though $C_{4}$ approaches a finite, nonzero value. This distinction also underlies the interpretation of the \textcolor{black}{flows with suppressed fermion fluctuations} in \cite{Guan:2026tbd}. {\color{black}At the numerical endpoint $t=-18$, we find $\Lambda^2C_4\simeq-\textcolor{black}{0.0503}$ and $-\textcolor{black}{0.00511}$ for $T=1.5\textcolor{black}{T_c}$ and $2\textcolor{black}{T_c}$, respectively, while $\tilde C_4$ is already of order $10^{-17}$ and $10^{-18}$.} {These values quantify how much of the generated mixed interaction remains once its flow nearly stops and show that its magnitude is smaller at the higher temperature. In the scalar-pseudoscalar basis and truncation used here, the mixed channel therefore contributes to the low-energy interaction even along regular thermal flows.}
	
	In Fig.~\ref{fig:frg-thermal-pole-scale}, we show how the divergence scale changes with temperature for the same UV conditions. At $T=0$, the couplings diverge at $k_{\rm pole}/\Lambda\simeq\textcolor{black}{0.808}$. As the temperature rises, this scale decreases, so the interactions must evolve further toward the IR before they diverge. The pole scale tends to zero as $T$ approaches $\textcolor{black}{T_c}$, and no finite-scale pole is found above this boundary. Consequently, the temperature effects can  prevent the strong-coupling instability even though the mixed interaction is generated along the flow.
	
	{For the benchmark in Figs.~\ref{fig:frg-generated-thermal-flows} and~\ref{fig:frg-thermal-pole-scale}, an axial rotation makes the mass purely scalar and diagonalizes the four-fermion interactions throughout the flow. The action therefore preserves parity in the rotated basis even as $\tilde C_4$ develops in the original basis.}
	
	\subsection{Connection with mean-field thermodynamics}
	\label{sec:mf-frg-connection}
	
	\begingroup\color{black}
	Both approaches associate the onset of instability with a vanishing inverse interaction. To show this connection, we consider $m=C_4=0$, where the scalar and pseudoscalar channels evolve independently. In mean-field theory, the relevant quantity is the second derivative of the potential at the origin. Equation~\eqref{eq:kernel} gives
	\begin{equation}
		a_i(T)\equiv
		\left.\frac{\partial^2\Omega_T}{\partial M_i^2}\right|_{M_S=M_P=0}
		=\frac{1}{G_i}-\EuScript{K}_3(0,T),
		\quad i=S,P.
		\label{eq:mf-frg-curvature}
	\end{equation}
	\textcolor{black}{A positive second derivative keeps the origin locally stable.} When it passes through zero and becomes negative, a condensate is favored in that channel.
	
	In the FRG, the corresponding quantity is the inverse dimensionful coupling. Integrating Eq.~\eqref{eq:frg-beta-functions} in the same limit gives
	\begin{equation}
		\begin{aligned}
			\frac{1}{G_{i,k}}&=\frac{1}{G_{i,\Lambda}}-\Pi_k(T),\\
			\Pi_k(T)&=8N_c\int_k^\Lambda\dd q\,q\,
			\widehat{\mathcal A}(T/q,0).
		\end{aligned}
		\label{eq:mf-frg-inverse-flow}
	\end{equation}
	{At fixed $T$, the fermion contribution $\Pi_k(T)$ accumulates as the flow runs from $\Lambda$ down to $k$, lowering the inverse coupling. If it reaches $1/G_{i,\Lambda}$ at a finite scale $k_{\rm pole}>0$, the inverse coupling vanishes and $G_{i,k}$ develops a pole, identifying the scale at which the fermionic flow encounters the instability. In the bosonized action, the inverse coupling sets the quadratic coefficient for forming a scalar or pseudoscalar field. The two calculations here use different cutoff and UV prescriptions, so their numerical critical couplings can differ. With matched prescriptions, the comparison at the thermal boundary uses $\Pi_0(T)$, accumulated by integrating the flow down to $k=0$. The condition $\Pi_0(T)=1/G_{i,\Lambda}$ then corresponds to the vanishing mean-field quadratic coefficient, where the origin loses stability~\cite{Braun:2011pp}.}
	\endgroup
	

	\subsection{Thermal evolution of the mass phase}
	\label{sec:frg-phase-evolution}
	
	We finally examine how temperature affects the running of the mass phase. The symmetric benchmark used above preserves $\theta=\pi/2$ and, consequently, does not give phase variation. To study a nontrivial response, we instead choose $\tilde G_S(0)=1.05G_*$, $\tilde G_P(0)=0$, $\tilde C_4(0)=0.15G_*$, and $\tilde m(0)=0.02$, with initial phases $\theta(0)=0$ and $\pi$. All five flow equations are retained. These trajectories remain regular over the explored interval.

	In Fig.~\ref{fig:frg-phase-evolution}, we show that the mass phase does not need to remain close to its UV value. It shows that the mixed interaction	induces a finite phase shift even for $\theta(0)=0$, while the vacuum trajectory starting at $\theta(0)=\pi$ undergoes a much larger change. Thus, for these UV interactions, these results show that $\theta$ participates in the coupled dynamics rather than 	acting only as an external parameter. This interaction-driven evolution is consistent with the mechanism discussed in \cite{Huang:2024ypj}, although the present regular	trajectories do not approach a universal limiting phase.
	{\color{black}Increasing the temperature reduces the accumulated shift and leaves the infrared phase closer to its initial value. The plateaus in Fig.~\ref{fig:frg-phase-evolution} show that the fluctuations driving the phase evolution become weak at low scales, leaving temperature-dependent infrared values.}
	
	\section{Conclusions and outlook}
	\label{sec:Conclusions}
	
	In this paper, we have investigated the thermal behavior of a one-flavor
	four-fermion model inspired by low-energy QCD, including
	scalar and pseudoscalar interactions, the $P$-odd mixed
	operator $(\bar\psi\psi)(\bar\psi i\gamma_5\psi)$, and a
	complex fermion mass. Motivated by the topological nature of the 
	$\theta$ term as a possible source of $\CP$ violation
	in QCD, our aim was to understand how temperature
	affects the interplay between fermionic interactions
	and explicit sources of $P/\CP$-violation. The model
	isolates these mechanisms without including dynamical
	gauge fields, and its thermal scales characterize
	the selected benchmarks rather than quantitative
	predictions for QCD. {Building on the vacuum studies~\cite{Huang:2024ypj,Guan:2026tbd}, we combine a finite-temperature FRG analysis with mean-field thermodynamics to distinguish symmetry restoration from thermal suppression in the presence of explicit sources.}
	
	At the mean-field level, we distinguished spontaneous
	symmetry restoration from thermal suppression of an
	explicitly induced response. In the source-free
	pseudoscalar regime, the parity-related minima merge
	continuously as the temperature increases, restoring
	ordinary $P/\CP$. Explicit sources, instead, generally produce
	smooth crossovers. {A purely pseudoscalar mass is an exception, since a residual scalar reflection can still undergo a continuous transition.} In particular, the mixed interaction
	induces a pseudoscalar condensate even when the
	pseudoscalar mass source vanishes. Hence, the decrease of
	the condensates does not need to imply the restoration
	of the symmetry of the action.
	
	For the FRG analysis, we employed a pointlike fermionic
	ansatz within the local potential approximation. The coupled flows show that the mixed interaction can be generated from a vanishing UV coefficient, including at a nonzero temperature.
	Increasing temperature weakens the growth of the
	interactions and, for the benchmarks studied,
	eventually removes their finite-scale divergence.
	{Along the regular high-temperature trajectories, the dimensionful mixed coefficient approaches a finite value as its flow nearly stops, while its dimensionless counterpart decreases through canonical scaling. Its low-scale value quantifies the interaction that remains at each temperature. These results support retaining the mixed operator $(\bar\psi\psi)(\bar\psi i\gamma_5\psi)$ in the scalar-pseudoscalar basis and pointlike truncation used here whenever the other couplings source its flow. In this setting, temperature can suppress the instability while leaving a finite mixed interaction.}
	
	The mass phase also participates in the coupled
	dynamics and can undergo an appreciable change,
	depending on the UV conditions. The higher temperature
	reduces the accumulated shift and leaves the
	infrared phase closer to its initial value.
	Its eventual freezing reflects suppression of
	the fluctuations responsible for its running,
	rather than evolution toward a universal
	symmetry-restoring value. Together, the mean-field
	and FRG results distinguish suppression of
	condensates and interaction growth from the elimination
	of explicit symmetry-breaking sources.
	
	{\color{black}A different infrared behavior has been proposed in a lattice study of Yang-Mills theory, where a scale-dependent topological angle approaches zero at long distances~\cite{Nakamura:2021}. In our model, instead, the running of $\theta$ describes the evolution of the fermion mass phase induced by the four-fermion interactions. Relating these two descriptions would require a treatment that includes the gauge sector and its matching to the fermionic effective theory.}
	
	{A natural next step is to follow the dynamics
		beyond the fermionic instability using a weak-solution formulation of the RG~\cite{Guan:2026weak} or dynamical hadronization (bosonization)~\cite{Gies:2002boundstates,Floerchinger:2009composites,Braun:2016dynamicalQCD}. This would let us follow the formation of the scalar and pseudoscalar condensates as the RG scale is lowered at fixed temperature. Repeating the calculation at different temperatures would give equilibrium condensate curves for direct comparison with mean-field theory.}
	
	Gauge-induced evolution of the mixed interaction
	has already been investigated in
	\cite{Guan:2026tbd}. Building on that work,
	a finite-temperature extension with a temporal
	gauge background and a gluonic Polyakov-loop
	potential would allow deconfinement effects and
	their interplay with chiral dynamics and
	$P/\CP$ violation to be explored. The potential
	would encode gluonic thermodynamics at an
	effective level rather than provide a
	first-principles derivation of confinement.
	
	Coupling the zero-temperature model with
	quantum metric fluctuations within asymptotically
	safe gravity would provide a distinct extension.
	Following the coupled evolution of $G_S$, $G_P$,
	and $C_4$, together with the mass parameters,
	would reveal whether gravitational fluctuations
	modify the fixed-point structure, enhance or
	suppress the mixed interaction, and alter the
	infrared evolution of existing $\CP$-violating
	sources.
	
	Finally, another possible application of the mixed type of interactions studied here would be to planar Dirac systems. Extending four-fermion descriptions of tilted Dirac materials~\cite{Gomes:2021nem} or excitonic insulators~\cite{Tavares:2026pio} by a mixed interaction could help explore competing gaps, thermal transitions, and Hall response. Such an application first requires identifying the bilinears with material order parameters and checking the allowed microscopic symmetries. We hope to address this and the other suggested applications and extensions in the future.

	\begingroup
	\begin{acknowledgments}
		
		A.F.V. is supported by FAPERJ through the P\'os-doutorado Nota 10
		program under Grants No.~200.135/2025 and No.~200.136/2025. R.O.R. is partially supported by research grants from Conselho
		Nacional de Desenvolvimento Cient\'{\i}fico e Tecnol\'ogico (CNPq),
		Grants No.~319307/2025-5 and No.~404964/2025-7, and from Funda\c{c}\~ao Carlos Chagas Filho
		de Amparo \`a Pesquisa do Estado do Rio de Janeiro (FAPERJ), Grant
		No. E-26/200.415/2026.
		{\color{black}OpenAI's Codex was used to assist with manuscript editing.}
		
	\end{acknowledgments}
	
	\section*{Data availability}
	The data supporting this study are available from the authors
	upon reasonable request.
	\endgroup
	
	\appendix
	
	\twocolumngrid
	
	\begingroup
	\section{Auxiliary fields and fermion integration}
	\label{app:meanfield-construction}
	
	The auxiliary-field formulation provides a direct route from the local four-fermion interactions to the thermodynamic potential. We first replace the quartic terms by a Gaussian integral, then integrate out the fermions for homogeneous auxiliary fields.
	
	It is convenient to collect the bilinears and their couplings as
	\begin{equation}
		\Jvec=\begin{pmatrix}\bar\psi\psi\\\bar\psi i\gamma_5\psi\end{pmatrix},\qquad
		\Gmat=\begin{pmatrix}G_S&C_4\\C_4&G_P\end{pmatrix},
		\label{eq:coupling-matrix}
	\end{equation}
	and introduce the auxiliary fields as $\bm\phi=(\sigma,\eta)^{\rm T}$. The conditions in Eq.~\eqref{eq:HS-domain} make $\Gmat$ positive definite. Consequently, the following Gaussian insertion is convergent,
	\begin{equation}
		1=\mathcal N\int[d\bm\phi]\,
		\exp\!\left[-\frac12\int_x
		(\bm\phi+\Gmat\Jvec)^{\rm T}\Gmat^{-1}
		(\bm\phi+\Gmat\Jvec)\right],
		\label{eq:HS-identity}
	\end{equation}
	where $\mathcal N$ is independent of the fields. Expanding the square produces the terms $-\bm\phi^{\rm T}\Gmat^{-1}\bm\phi/2$, $-\bm\phi^{\rm T}\Jvec$, and $-\Jvec^{\rm T}\Gmat\Jvec/2$. The last term cancels the original four-fermion interaction in $e^{-S_{\rm E}}$, leaving the bosonized action in Eq.~\eqref{eq:bosonized-action}. This change of variables is exact.
	
	We now take $\sigma$ and $\eta$ to be homogeneous and integrate over the fermions. At a fermionic Matsubara frequency $\omega_n=(2n+1)\pi T$, the Dirac operator is
	\begin{equation}
		\mathcal D_n(\bm p)=i\gamma_0\omega_n+i\bm\gamma\cdot\bm p
		+M_S+i\gamma_5M_P,
		\label{eq:thermal-Dirac-operator}
	\end{equation}
	where mass components are $M_S=m_S+\bar\sigma$ and $M_P=m_P+\bar\eta$, and $M^2=M_S^2+M_P^2$.
	{\color{black}Using the Euclidean gamma-matrix algebra, its determinant over spinor indices is}
	\begin{equation}
		\Det\mathcal D_n(\bm p)
		=(\omega_n^2+p^2+M^2)^2.
		\label{eq:spinor-determinant}
	\end{equation}
	The fermion contribution therefore depends only on the total mass $M$, even when both condensates are present. {\color{black}Integrating out the fermions gives minus the logarithm of this determinant, summed over colors and fermionic modes.} Defining $E_p=\sqrt{p^2+M^2}$ and $\int_{\bm p}=\int d^3p/(2\pi)^3$, we obtain
	\begin{equation}
		\begin{aligned}
			F_T(M)&=\textcolor{black}{-N_cT\sum_n\int_{\bm p}
				\ln\Det\mathcal D_n(\bm p)}\\
			&=-2N_cT\sum_n\int_{\bm p}
			\ln(\omega_n^2+E_p^2).
		\end{aligned}
		\label{eq:fermion-determinant}
	\end{equation}
	The sum gives $E_p+2T\ln(1+e^{-E_p/T})$, up to a field-independent constant. Angular integration with $\vb*{p}\leq\Lambda_3$ then yields Eq.~\eqref{eq:fermion-loop}. Together with the quadratic auxiliary-field term, this gives the potential in Eq.~\eqref{eq:Omega-bar}. The normalization follows from $\Omega_T=-(T/V)\ln\mathcal Z_{\rm MF}$.
	
	The condensates follow from the same construction. A shift of the auxiliary integration variables leaves its measure unchanged and gives $\langle\Gmat^{-1}\bm\phi+\Jvec\rangle_T=0$. At the homogeneous saddle this becomes
	\begin{equation}
		\begin{pmatrix}\langle\bar\psi\psi\rangle_T\\
			\langle\bar\psi i\gamma_5\psi\rangle_T\end{pmatrix}
		=-\Gmat^{-1}\begin{pmatrix}\bar\sigma\\\bar\eta\end{pmatrix}.
		\label{eq:auxiliary-condensate-relation}
	\end{equation}
	Thus the auxiliary fields and condensates are related by the full interaction matrix.
	\endgroup
	
	\begingroup\color{black}
	\setlength{\parskip}{0pt}
	\section{Analytic properties of the thermal kernel}
	\label{app:kernel}
	
	In this appendix, we collect the properties of the mean-field fermion-loop function $\EuScript{K}_3(M,T)$, defined in Eq.~\eqref{eq:kernel}, which explain the critical coupling and the temperature dependence of the condensates. We use $\Lambda_3=\Lambda/\sqrt2$.
	
	\subsection{Vacuum critical coupling and temperature dependence}
	
	{\color{black}For convenience, we recall the kernel in Eq.~\eqref{eq:kernel},}
	\begin{equation}
		\textcolor{black}{\EuScript{K}_3(M,T)=\frac{N_c}{\pi^2}
			\int_0^{\Lambda_3}\!dp\,\frac{p^2}{E_p}
			\tanh\!\left(\frac{E_p}{2T}\right)}
		\label{eq:appendix-thermal-kernel}
	\end{equation}
	and we choose to use the same spatial cutoff for the vacuum and thermal integrals\footnote{\color{black}In the Medium Separation Scheme, UV-divergent vacuum contributions are separated from the medium-dependent terms~\cite{Farias:2016let}. The choice of prescription can affect the thermal response. Here, both vacuum and thermal terms use the cutoff in Eq.~\eqref{eq:fermion-loop}.}. At zero mass and temperature,
	\begin{equation}
		\EuScript{K}_3(0,0)=\frac{N_c\Lambda_3^2}{2\pi^2}
		=\frac{N_c\Lambda^2}{4\pi^2}.
		\label{eq:cutoff-matching}
	\end{equation}
	{\color{black}This choice gives the same critical coupling as an $O(4)$-symmetric sharp cutoff $\Lambda$. The full potentials obtained with the two prescriptions can differ.} The critical coupling in Fig.~\ref{fig:zeroT-phase-map} is
	\begin{equation}
		G_i=g_c=\frac{1}{\EuScript{K}_3(0,0)},
		\qquad i=S,P.
		\label{eq:kernel-critical-coupling}
	\end{equation}
	For nonzero mass, direct integration gives
	\begin{equation}
		\begin{aligned}
			\EuScript{K}_3(M,0)=\frac{N_c}{2\pi^2}\bigg[
			&\Lambda_3\sqrt{\Lambda_3^2+M^2}-M^2\operatorname{arsinh}\!\frac{\Lambda_3}{M}\bigg].
		\end{aligned}
		\label{eq:vacuum-kernel}
	\end{equation}
	Its massless limit recovers Eq.~\eqref{eq:cutoff-matching}. At finite temperature, the massless expression becomes
	\begin{equation}
		\EuScript{K}_3(0,T)=\frac{N_c}{\pi^2}
		\left[\frac{\Lambda_3^2}{2}
		-2T^2\int_0^{\Lambda_3/T}\frac{u\,du}{e^u+1}\right].
		\label{eq:origin-kernel}
	\end{equation}
	The negative second term shows that increasing temperature decreases $\EuScript{K}_3(0,T)$. For $T\ll\Lambda_3$, the integral approaches $\pi^2/12$, yielding
	\begin{equation}
		\EuScript{K}_3(0,T)\simeq
		\frac{N_c\Lambda^2}{4\pi^2}-\frac{N_cT^2}{6},
		\label{eq:kernel-low-temperature}
	\end{equation}
	with exponentially small cutoff corrections.
	
	\subsection{High-temperature limit}
	\label{app:high-temperature-limit}
	
	At fixed cutoff and couplings, the gap equations (cf. Eqs.\eqref{eq:gap-scalar} and \eqref{eq:gap-pseudoscalar}) take the form
	\begin{equation}
		\begin{aligned}
			\bar\sigma&=\EuScript{K}_3\,(G_SM_S+C_4M_P),\\
			\bar\eta&=\EuScript{K}_3\,(C_4M_S+G_PM_P).
		\end{aligned}
		\label{eq:kernel-field-gap}
	\end{equation}
	The high temperature limit means $T\gg\sqrt{\Lambda_3^2+M^2}$, so all fermion energies included in the integral are small compared with $T$. Expanding the hyperbolic tangent in Eq.~\eqref{eq:kernel} then gives
	\begin{equation}
		\begin{aligned}
			\EuScript{K}_3(M,T)={}&\frac{N_c\Lambda_3^3}{6\pi^2T}+\textcolor{black}{O(T^{-3})}.
		\end{aligned}
		\label{eq:highT-kernel-derivation}
	\end{equation}
	Thus $\EuScript{K}_3$ vanishes, and the finite solution approaches
	\begin{equation}
		\begin{aligned}
			(\bar\sigma,\bar\eta)&\longrightarrow(0,0),\\
			(M_S,M_P)&\longrightarrow(m_S,m_P).
		\end{aligned}
		\label{eq:highT-mass-limit}
	\end{equation}
	Both fermion condensates therefore vanish as temperature increases, suppressing the condensate response while leaving the explicit mass sources and any prescribed mixed interaction unaffected.
	
	\subsection{Monotonicity and the continuous transition}
	
	{\color{black}Differentiating the kernel integral in Eq.~\eqref{eq:kernel} with respect to $M^2$, at fixed $T>0$ and cutoff, gives
		\begin{equation}
			\frac{\partial\EuScript{K}_3}{\partial M^2}
			=-\frac{N_c}{2\pi^2}\int_0^{\Lambda_3}dp\,
			\frac{p^2}{E_p^3}\,
			f\!\left(\frac{E_p}{2T}\right),
			\label{eq:Kx-explicit}
		\end{equation}
		where
		\begin{equation}
			\begin{aligned}
				f(u)&=\tanh u-u\operatorname{sech}^2u,\\
				f'(u)&=2u\operatorname{sech}^2u\tanh u>0
				\quad (u>0).
			\end{aligned}
			\label{eq:kernel-monotonicity-function}
		\end{equation}
		Since $f(0)=0$ and $f'(u)>0$ for $u>0$, the derivative in Eq.~\eqref{eq:Kx-explicit} is negative. Thus, $\EuScript{K}_3$ decreases with $M^2$.} In this way, the quartic coefficient of the mean-field potential in Eq.~\eqref{eq:landau-expansion} is positive,
	\begin{equation}
		b(T)=-\left.\frac{\partial\EuScript{K}_3}{\partial M^2}\right|_{M=0}>0.
		\label{eq:kernel-positive-quartic}
	\end{equation}
	The source-free condensate then vanishes continuously when the \textcolor{black}{quadratic coefficient of the potential} reaches zero. The vacuum expression in Eq.~\eqref{eq:vacuum-kernel} also decreases with $M>0$.
	
	\subsection{Residual scalar-reflection transition}
	
	For a purely pseudoscalar mass source, $m_S=C_4=0$ and $m_P=m$. Then $M_S=\bar\sigma$ and $M_P=m+\bar\eta$, and the gap equations reduce to
	\begin{align}
		M_S[1-G_S\EuScript{K}_3(M,T)]&=0,
		\label{eq:residual-scalar-gap}\\
		M_P-m&=G_P\EuScript{K}_3(M,T)M_P.
		\label{eq:residual-pseudoscalar-gap}
	\end{align}
	For a solution with $M_S\ne0$, the first equation requires $\EuScript{K}_3=1/G_S$. Substituting this into the second gives
	\begin{equation}
		M_P=\frac{G_Sm}{G_S-G_P}.
		\label{eq:residual-fixed-mass}
	\end{equation}
	The pseudoscalar condensate is consequently
	\begin{equation}
		\langle\bar\psi i\gamma_5\psi\rangle_T
		=-\EuScript{K}_3M_P=-\frac{m}{G_S-G_P}.
		\label{eq:residual-condensate-plateau}
	\end{equation}
	It remains constant while the scalar component is nonzero, explaining the plateau in last panel of Fig.~\ref{fig:mass-phase}. The two minima have opposite scalar components and the same fixed pseudoscalar component. At their merger, $M_S=0$ and $M=|M_P|$, so
	\begin{equation}
		\EuScript{K}_3\!\left(\left|\frac{G_Sm}{G_S-G_P}\right|,T_{\rm r}\right)
		=\frac1{G_S}.
		\label{eq:residual-temperature}
	\end{equation}
	For the benchmark in Eq.~\eqref{eq:mass-benchmark}, $M_P/\Lambda=0.30$, $R_P=0.20$, and $T_{\rm r}/\Lambda\simeq0.235$. Above $T_{\rm r}$, the scalar gap equation is satisfied by $M_S=0$ and no longer fixes $\EuScript{K}_3=1/G_S$. The pseudoscalar condensate can then decrease. {\color{black}As Fig.~\ref{fig:residual-contours} shows, $T_{\rm r}$ marks the scalar-reflection transition.}
	\endgroup
	
	\twocolumngrid
	
	\begingroup
	\section{Derivation of the RG-flow equations}
	\label{app:frg-derivation}
	\allowdisplaybreaks[0]
	\makeatletter
	\patchcmd{\widetext}{\leaders\hrule\hfil}{\hfil}{}{}
	\patchcmd{\widetext}{\leaders\hrule\hfil}{\hfil}{}{}
	\patchcmd{\widetext}{\vrule\@height6\p@}{\vrule\@width0pt\@height6\p@}{}{}
	\patchcmd{\widetext}{\vrule\@depth6\p@}{\vrule\@width0pt\@depth6\p@}{}{}
	\makeatother
	
	The flow equations in Sec.~\ref{sec:frg-flows} describe how fermion fluctuations modify the mass, its phase, and the three four-fermion interactions. This appendix explains their derivation from the truncated effective action. We use the Euclidean conventions of Appendix A of \cite{Huang:2024ypj}, with $\{\gamma_\mu,\gamma_\nu\}=2\delta_{\mu\nu}$, $\{\gamma_5,\gamma_\mu\}=0$, $\gamma_5^2=1$, and $\operatorname{tr}_D(\gamma_\mu\gamma_\nu)=4\delta_{\mu\nu}$. There is one flavor throughout, $N_f=1$.
	
	At fixed physical temperature, we adopt the following shorthand notation for each loop integral:
	\begin{equation}
		\int_p^{(T)}\equiv T\sum_{n\in\mathbb Z}
		\int\frac{d^3\bm p}{(2\pi)^3},\qquad
		p_0=\omega_n=(2n+1)\pi T.
		\label{eq:frg-matsubara}
	\end{equation}
	The integer $n$ labels the fermionic Matsubara frequencies, while the spatial momentum remains continuous. With this measure, the momentum-space identity is
	\begin{equation}
		\delta_{p,q}=\beta(2\pi)^3\delta_{n_p,n_q}
		\delta^{(3)}(\bm p-\bm q),\qquad \beta=1/T.
		\label{eq:frg-momentum-delta}
	\end{equation}
	Thus, $\int_q^{(T)}\delta_{p,q}f(q)=f(p)$. We suppress the scale label on $m$, $\theta$, $G_S$, $G_P$, and $C_4$ when no ambiguity arises.
	
	\subsection{Fermion Hessian and interaction vertices}
	
	The second derivatives of the effective action are conveniently arranged using the superfield $\Phi=(\psi,\bar\psi^{\,T})^{\rm T}$. With the Grassmann derivative ordering fixed by
	\begin{equation}
		(\Gamma_k^{(2)})_{IJ}(p,q)=
		\frac{\overrightarrow\delta}{\delta\Phi_I^T(-p)}\Gamma_k
		\frac{\overleftarrow\delta}{\delta\Phi_J(q)},
		\label{eq:frg-hessian}
	\end{equation}
	the fermionic Hessian separates as
    \begin{equation}
        \Gamma_k^{(2)}(p,q)=\mathcal K_k(p,q)+\mathcal V_k(p,q).
    \end{equation}
    Its field-independent part is
	\begin{equation}
		\mathcal K_k(p,q)=\delta_{p,q}
		\begin{pmatrix}
			0&(i\gamma\cdot p-me^{i\gamma_5\theta/2})^{T}\\
			i\gamma\cdot p+me^{i\gamma_5\theta/2}&0
		\end{pmatrix}.
		\label{eq:frg-kernel}
	\end{equation}
	Before taking the large-$N_c$ limit, the interaction matrix contains four blocks,
	\begin{equation}
		\mathcal V_k(p,q)=\int_x e^{-i(q-p)\cdot x}
		\begin{pmatrix}V_{11}&V_{12}\\V_{21}&V_{22}\end{pmatrix}_x.
		\label{eq:frg-vertex-matrix}
	\end{equation}
	To capture these contributions explicitly, we introduce the local Dirac matrix $\mathcal V_{\rm loc}$, which combines the scalar and pseudoscalar channels as follows:
	\begin{equation}
		\begin{aligned}
			\mathcal V_{\rm loc}&=b_S\bm 1+b_Pi\gamma_5,\\
			b_S&=G_S(\bar\psi\psi)+C_4(\bar\psi i\gamma_5\psi),\\
			b_P&=C_4(\bar\psi\psi)+G_P(\bar\psi i\gamma_5\psi).
		\end{aligned}
		\label{eq:frg-Bvertex}
	\end{equation}
	Writing the spinor and color indices as $i,j$, the four blocks are
	\begin{align}
		\begin{split}
			(V_{11})_{ij}={}&G_S\bar\psi_i\bar\psi_j
			+G_P(\bar\psi i\gamma_5)_i(\bar\psi i\gamma_5)_j\\
			&+C_4\big[\bar\psi_i(\bar\psi i\gamma_5)_j+(\bar\psi i\gamma_5)_i\bar\psi_j\big],
		\end{split}
		\label{eq:frg-vertex-11}\\
		\begin{split}
			(V_{22})_{ij}={}&G_S\psi_i\psi_j
			+G_P(i\gamma_5\psi)_i(i\gamma_5\psi)_j\\
			&+C_4\big[\psi_i(i\gamma_5\psi)_j+(i\gamma_5\psi)_i\psi_j\big],
		\end{split}
		\label{eq:frg-vertex-22}\\
		\begin{split}
			(V_{12})_{ij}={}&-G_S\bar\psi_i\psi_j
			-G_P(\bar\psi i\gamma_5)_i(i\gamma_5\psi)_j\\
			&-C_4\big[\bar\psi_i(i\gamma_5\psi)_j+(\bar\psi i\gamma_5)_i\psi_j\big]
			+(\mathcal V_{\rm loc}^{T})_{ij},
		\end{split}
		\label{eq:frg-vertex-12}\\
		\begin{split}
			(V_{21})_{ij}={}&-G_S\psi_i\bar\psi_j
			-G_P(i\gamma_5\psi)_i(\bar\psi i\gamma_5)_j\\
			&-C_4\big[\psi_i(\bar\psi i\gamma_5)_j+(i\gamma_5\psi)_i\bar\psi_j\big]
			-(\mathcal V_{\rm loc})_{ij}.
		\end{split}
		\label{eq:frg-vertex-21}
	\end{align}
	These expressions keep the Grassmann-field ordering explicit. The matrices $\bm 1$ and $i\gamma_5$ act trivially in color space. In the projections below, the terms containing $\mathcal V_{\rm loc}$ produce a closed color trace and hence a factor $N_c$. The remaining terms are exchange contributions that are subleading in these projections. In the large-$N_c$ approximation, we retain this leading color enhancement. We can then drop $V_{11}$ and $V_{22}$ entirely, together with all terms in $V_{12}$ and $V_{21}$ other than $\mathcal V_{\rm loc}^{T}$ and $-\mathcal V_{\rm loc}$, respectively. The retained vertex matrix is then
	\begin{equation}
		\begin{aligned}
			\mathcal V_k(p,q)\simeq{}&\int_x e^{-i(q-p)\cdot x}
			\begin{pmatrix}0&\mathcal V_{\rm loc}^{T}\\
				-\mathcal V_{\rm loc}&0\end{pmatrix}_x.
		\end{aligned}
		\label{eq:frg-leading-vertex}
	\end{equation}
	For a homogeneous background, the Fourier integral in Eq.~\eqref{eq:frg-vertex-matrix} reduces to $\delta_{p,q}$ times the local block matrix, with $\delta_{p,q}$ defined in Eq.~\eqref{eq:frg-momentum-delta}. For a general background, the Fourier form retains the momentum transfer carried by the external fields. Each block is quadratic in these fields, and $C_4$ contributes to both the scalar and pseudoscalar parts.
	
	\subsection{Regulated propagator and vertex expansion}
	
	We use the same spatial Litim regulator as in the main text,
	\begin{equation}
		\begin{aligned}
			\mathcal R_k(p,q)&=\delta_{p,q}
			\begin{pmatrix}0&i\slashed{\bm p}^{\,T}r_F\\
				i\slashed{\bm p}\,r_F&0\end{pmatrix},\\
			r_F&=\left(\sqrt{\frac{k^2}{\bm p^2}}-1\right)
			\Theta(k^2-\bm p^2).
		\end{aligned}
		\label{eq:frg-appendix-regulator}
	\end{equation}
	It acts on spatial momenta while leaving $\omega_n$ unchanged. Introducing $\bar p=(\omega_n,(1+r_F)\bm p)$ and $D_n=\bar p^2+m^2$, the field-independent regulated fermion propagator $\mathcal P_k=(\mathcal K_k+\mathcal R_k)^{-1}$ is explicitly
	\begin{equation}
		\begin{aligned}
			\mathcal P_k(p,q)
			&=\delta_{p,q}
			\begin{pmatrix}0&P_k(p)\\-P_k^T(-p)&0\end{pmatrix}.
			\\
			P_k(p)&=\frac{-i\gamma\cdot\bar p+me^{-i\gamma_5\theta/2}}{D_n}.
		\end{aligned}
		\label{eq:frg-propagator}
	\end{equation}
	After factoring out $\delta_{p,q}$, this equation also defines the Dirac propagator $P_k$ used in the projections below. For the chosen regulator, the denominator inside the spatial integration region $|\bm p|<k$ is simply $D_n=\omega_n^2+k^2+m^2$.
	
	The fermionic Wetterich equation can be expressed in terms of the regulated propagator $\mathcal P_k$ and expanded in powers of $\mathcal V_k$. Through fourth order in the external fields, it reads
	\begin{widetext}
		\begin{equation}
			\begin{aligned}
				\partial_t\Gamma_k&=\frac12\,\mathrm{STr}\left[
				\left(\Gamma_k^{(2)}+ \mathcal R_k\right)^{-1}\cdot\partial_t \mathcal R_k\right]
				=-\frac{1}{2}\tilde\partial_t\operatorname{Tr}\ln(\mathcal P_k^{-1}+\mathcal V_k)\\
				&=-\frac12\operatorname{Tr}
				(\mathcal P_k\partial_t\mathcal R_k)
				+\frac12\operatorname{Tr}
				(\mathcal P_k\mathcal V_k\mathcal P_k\partial_t\mathcal R_k)-\frac12\operatorname{Tr}
				(\mathcal P_k\mathcal V_k\mathcal P_k\mathcal V_k
				\mathcal P_k\partial_t\mathcal R_k)
				+O(\mathcal V_k^3).
			\end{aligned}
			\label{eq:frg-vertex-expansion}
		\end{equation}
	\end{widetext}
	Here $\operatorname{Tr}$ denotes the trace over momentum, Dirac, color, and doubled-field indices and $\tilde\partial_t$ acts only on the regulator. In particular, $\tilde\partial_t\mathcal P_k=-\mathcal P_k(\partial_t\mathcal R_k)\mathcal P_k$ and $\tilde\partial_t\mathcal V_k=0$. The first term is independent of the external fermions and is not needed for the couplings considered here.
	
	\subsection{Two-point projection: mass and phase couplings}
	
	{\color{black}The term linear in $\mathcal V_k$ in Eq.~\eqref{eq:frg-vertex-expansion} contains two external fermions. The two off-diagonal blocks give equal contributions, canceling its factor $1/2$. Their flow then becomes}
	\begingroup
	\setlength{\abovedisplayskip}{6pt}
	\setlength{\abovedisplayshortskip}{6pt}
	\begin{equation}
		\left.\partial_t\Gamma_k\right|_{2\mathrm{pt}}
		=-N_c\int_x\int_p^{(T)}\partial_t r_F\times\operatorname{tr}_D
		\left[P_k\mathcal V_{\rm loc}P_k\,i\bm\gamma\cdot\bm p\right].
		\label{eq:frg-two-point-start}
	\end{equation}
	\endgroup
	The color trace supplies $N_c$, while $\operatorname{tr}_D$ now acts only on spinor indices. To evaluate this trace, write the propagator numerator as $K+mU$, with $K=-i\gamma\cdot\bar p$ and $U=e^{-i\gamma_5\theta/2}$. Its two factors give
	\begin{equation}
		\begin{aligned}
			&(K+mU)\mathcal V_{\rm loc}(K+mU)\\
			&\quad=K\mathcal V_{\rm loc}K
			+m(U\mathcal V_{\rm loc}K+K\mathcal V_{\rm loc}U)+m^2U\mathcal V_{\rm loc}U.
		\end{aligned}
		\label{eq:frg-two-point-numerator}
	\end{equation}
	After multiplication by the regulator insertion, the first and last terms have vanishing Dirac traces. The two cross terms survive. Using the anticommutation of $\gamma_5$ with $\gamma_\mu$ gives
	\begin{align}
		\begin{split}
			\operatorname{tr}_D(U\mathcal V_{\rm loc})
			=4\big[b_S\cos(\theta/2)+b_P\sin(\theta/2)\big],
		\end{split}
		\label{eq:frg-two-point-trace-U}\\
		\begin{split}
			&\operatorname{tr}_D[(U\mathcal V_{\rm loc}K
			+K\mathcal V_{\rm loc}U)i\bm\gamma\cdot\bm p]\\
			&\quad=8(1+r_F)\bm p^2
			\big[b_S\cos(\theta/2)+b_P\sin(\theta/2)\big].
		\end{split}
		\label{eq:frg-two-point-trace}
	\end{align}
	Thus the matrix expression in Eq.~\eqref{eq:frg-two-point-start} reduces to a scalar loop integral multiplying the two bilinears,
	\begin{widetext}
		\begin{equation}
			\begin{aligned}
				\left.\partial_t\Gamma_k\right|_{2\mathrm{pt}}
				&=-8N_cm\int_x\int_p^{(T)}
				\frac{\bm p^2(1+r_F)\partial_t r_F}{D_n^2}
				\big[b_S\cos(\theta/2)+b_P\sin(\theta/2)\big]\\
				&=-8N_cm\mathcal M\int_x
				\Big\{\big[G_S\cos(\theta/2)+C_4\sin(\theta/2)\big](\bar\psi\psi)
				+\big[C_4\cos(\theta/2)+G_P\sin(\theta/2)\big]
				(\bar\psi i\gamma_5\psi)\Big\},
			\end{aligned}
			\label{eq:frg-two-point-result}
		\end{equation}
	\end{widetext}
	where the common threshold function is
	\begin{equation}
		\mathcal M=\int_p^{(T)}
		\frac{\bm p^2(1+r_F)\partial_t r_F}{D_n^2}.
		\label{eq:frg-M}
	\end{equation}
	The second line of Eq.~\eqref{eq:frg-two-point-result} follows by inserting $b_S$ and $b_P$ from Eq.~\eqref{eq:frg-Bvertex}. Matching the coefficients of $\bar\psi\psi$ and $\bar\psi i\gamma_5\psi$ to the mass terms in the action then gives:
	\begin{align}
		\begin{split}
			\partial_t[m\cos(\theta/2)]
			={}&-8N_cm\big[G_S\cos(\theta/2)\\
			&\qquad+C_4\sin(\theta/2)\big]\mathcal M,
		\end{split}
		\label{eq:frg-component-scalar}\\
		\begin{split}
			\partial_t[m\sin(\theta/2)]
			={}&-8N_cm\big[C_4\cos(\theta/2)\\
			&\qquad+G_P\sin(\theta/2)\big]\mathcal M.
		\end{split}
		\label{eq:frg-component-pseudoscalar}
	\end{align}
	Multiplying Eq.~\eqref{eq:frg-component-scalar} by $\cos(\theta/2)$ and Eq.~\eqref{eq:frg-component-pseudoscalar} by $\sin(\theta/2)$, then adding the two equations, gives the mass flow,
	\begin{equation}
		\begin{aligned}
			\partial_tm={}&\cos(\theta/2)\partial_t[m\cos(\theta/2)]\\
			&+\sin(\theta/2)\partial_t[m\sin(\theta/2)].
		\end{aligned}
		\label{eq:frg-radial-projection}
	\end{equation}
	The phase follows by comparing the $i\gamma_5$ coefficient in
	$e^{-i\gamma_5\theta/2}\partial_t(me^{i\gamma_5\theta/2})=\partial_tm+i\gamma_5m\partial_t\theta/2$. Using $\tilde m=m/k$, $\tilde G_i=k^2G_i$, $\tilde C_4=k^2C_4$, and $\widehat{\mathcal M}=\mathcal M/k^2$, we then obtain
	\begin{align}
		\begin{split}
			\partial_t\tilde m={}&-\tilde m-4N_c\tilde m\big[
			(\tilde G_S+\tilde G_P)\\
			&+(\tilde G_S-\tilde G_P)\cos\theta
			+2\tilde C_4\sin\theta\big]\widehat{\mathcal M},
		\end{split}
		\label{eq:frg-appendix-beta-m}\\
		\begin{split}
			\partial_t\theta={}&8N_c\widehat{\mathcal M}\big[
			(\tilde G_S-\tilde G_P)\sin\theta\\
			&\hphantom{8N_c\widehat{\mathcal M}\big[}
			-2\tilde C_4\cos\theta\big].
		\end{split}
		\label{eq:frg-appendix-beta-theta}
	\end{align}
	In the massless limit, both RG-flow equations remain well defined and preserve zero mass. Since the mass phase is undefined in this limit, no separate value of $\theta$ is needed.
	\subsection{Four-point projection: four-fermion couplings}
	
	{\color{black}The term quadratic in $\mathcal V_k$ in Eq.~\eqref{eq:frg-vertex-expansion} contains four external fermions. After taking the traces over the superfield and color indices, we obtain}
	\begin{equation}
		\begin{aligned}
			\left.\partial_t\Gamma_k\right|_{4\mathrm f}
			={}&-N_c\int_x\int_p^{(T)}\partial_t r_F\\
			&\times\operatorname{tr}_D\big[
			P_k\mathcal V_{\rm loc}P_k\mathcal V_{\rm loc}P_k\,i\bm\gamma\cdot\bm p\big].
		\end{aligned}
		\label{eq:frg-four-point-start}
	\end{equation}
	Its three propagator numerators produce eight terms. Those with one or three mass factors vanish under the Dirac trace. The remaining contributions contain either no mass factor or two mass factors,
	\begin{widetext}
		\begin{gather}
			\operatorname{tr}_D
			[K\mathcal V_{\rm loc}K\mathcal V_{\rm loc}K\,i\bm\gamma\cdot\bm p]
			=-4(1+r_F)\bm p^2(b_S^2+b_P^2)\bar p^2,
			\label{eq:frg-four-point-trace-kinetic}\\*[6pt]
			\begin{aligned}
				m^2\operatorname{tr}_D\big[
				(U\mathcal V_{\rm loc}U\mathcal V_{\rm loc}K
				+U\mathcal V_{\rm loc}K\mathcal V_{\rm loc}U
				+K\mathcal V_{\rm loc}U\mathcal V_{\rm loc}U)
				i\bm\gamma\cdot\bm p\big]\\
				=4(1+r_F)\bm p^2m^2
				\big[b_S^2+b_P^2+2(b_S^2-b_P^2)\cos\theta
				+4b_Sb_P\sin\theta\big].
			\end{aligned}
			\label{eq:frg-four-point-trace-mass}
		\end{gather}
	\end{widetext}
	Substituting these traces into Eq.~\eqref{eq:frg-four-point-start} combines the momentum-dependent terms into
	\begin{equation}
		\begin{aligned}
			\left.\partial_t\Gamma_k\right|_{4\mathrm f}
			&=4N_c\int_x\int_p^{(T)}
			\frac{\bm p^2(1+r_F)\partial_t r_F}{D_n^3}\\
			&\times\Big\{(b_S^2+b_P^2)(\bar p^2-m^2)\\
			&-2m^2\big[(b_S^2-b_P^2)\cos\theta+2b_Sb_P\sin\theta\big]\Big\}.
		\end{aligned}
		\label{eq:frg-four-point-result}
	\end{equation}
	This expression still contains the external fields through $b_S$ and $b_P$. Expanding their products makes the three interaction channels explicit,
	\begin{equation}
		\begin{aligned}
			b_S^2+b_P^2={}&(G_S^2+C_4^2)(\bar\psi\psi)^2\\
			&+(G_P^2+C_4^2)(\bar\psi i\gamma_5\psi)^2\\
			&+2(G_S+G_P)C_4(\bar\psi\psi)(\bar\psi i\gamma_5\psi),
		\end{aligned}
		\label{eq:frg-products-sum}
	\end{equation}
	\begin{equation}
		\begin{aligned}
			b_S^2-b_P^2={}&(G_S^2-C_4^2)(\bar\psi\psi)^2\\
			&-(G_P^2-C_4^2)(\bar\psi i\gamma_5\psi)^2\\
			&+2(G_S-G_P)C_4(\bar\psi\psi)(\bar\psi i\gamma_5\psi),
		\end{aligned}
		\label{eq:frg-products-difference}
	\end{equation}
	\begin{equation}
		\begin{aligned}
			b_Sb_P={}&G_SC_4(\bar\psi\psi)^2+G_PC_4(\bar\psi i\gamma_5\psi)^2\\
			&+(G_SG_P+C_4^2)(\bar\psi\psi)(\bar\psi i\gamma_5\psi).
		\end{aligned}
		\label{eq:frg-products-mixed}
	\end{equation}
	The loop integrals multiplying these products can be organized into two threshold functions,
	\begin{align}
		\mathcal A&=\int_p^{(T)}\bm p^2(1+r_F)\partial_t r_F
		\left(\frac1{D_n^2}-\frac{2m^2}{D_n^3}\right),
		\label{eq:frg-A}\\
		\mathcal B&=\int_p^{(T)}\frac{\bm p^2(1+r_F)\partial_t r_F}{D_n^3}.
		\label{eq:frg-B}
	\end{align}
	In particular, $\bar p^2-m^2=D_n-2m^2$ explains the two terms in $\mathcal A$. The remaining mass-dependent terms in Eq.~\eqref{eq:frg-four-point-result} share $\mathcal B$.
	
	We now read off the coefficients of $(\bar\psi\psi)^2$, $(\bar\psi i\gamma_5\psi)^2$, and $(\bar\psi\psi)(\bar\psi i\gamma_5\psi)$. The action contains these operators with coefficients $-G_S/2$, $-G_P/2$, and $-C_4$, respectively. The corresponding beta functions are
	\begin{widetext}
		\begin{align}
			\partial_tG_S={}&-8N_c\Big[(G_S^2+C_4^2)\mathcal A
			-2m^2(G_S^2-C_4^2)\cos\theta\,\mathcal B
			-4m^2G_SC_4\sin\theta\,\mathcal B\Big],
			\label{eq:frg-appendix-beta-GS}\\*
			\partial_tG_P={}&-8N_c\Big[(G_P^2+C_4^2)\mathcal A
			+2m^2(G_P^2-C_4^2)\cos\theta\,\mathcal B
			-4m^2G_PC_4\sin\theta\,\mathcal B\Big],
			\label{eq:frg-appendix-beta-GP}\\*
			\partial_tC_4={}&-8N_c\Big[(G_S+G_P)C_4\mathcal A
			-2m^2(G_S-G_P)C_4\cos\theta\,\mathcal B
			-2m^2(G_SG_P+C_4^2)\sin\theta\,\mathcal B\Big].
			\label{eq:frg-appendix-beta-C4}
		\end{align}
	\end{widetext}
	The dimensionless equations in the main text follow by differentiating $k^2G_i$ and $k^2C_4$. Their canonical terms are therefore $2\tilde G_i$ and $2\tilde C_4$. The loop terms use $\widehat{\mathcal A}=\mathcal A/k^2$ and $\widehat{\mathcal B}=\mathcal B$, together with the threshold functions evaluated below. The channel structure agrees with the purely fermionic projection in~\cite{Guan:2026tbd}, with the spatial cutoff and thermal frequency sum used here.
	
	\subsection{Thermal threshold functions}
	
	\setlength{\abovedisplayskip}{6pt}
	\setlength{\belowdisplayskip}{6pt}
	
	The spatial Litim regulator makes the denominators independent of $|\bm p|$ inside $|\bm p|<k$. Its insertion supplies a factor $k^2$, and integration over this ball supplies $k^3/(6\pi^2)$. Thus all three threshold functions reduce to frequency sums at the energy $E=\sqrt{k^2+m^2}$. They can be evaluated from
	\begin{align}
		S_1(E,T)&\equiv T\sum_n\frac1{\omega_n^2+E^2}
		=\frac{\tanh(E/2T)}{2E},\\
		S_{j+1}&=-\frac1{2jE}\frac{\partial S_j}{\partial E}.
	\end{align}
	\label{eq:frg-matsubara-recursion}
	The derivative is taken at fixed $T$. The sums required in the projections are
	\begin{align}
		S_2(E,T)&=\frac{\tanh(E/2T)}{4E^3}
		-\frac{\operatorname{sech}^2(E/2T)}{8TE^2},
		\label{eq:frg-sum-S2}\\
		\begin{split}
			S_3(E,T)={}&\frac{3\tanh(E/2T)}{16E^5}
			-\frac{3\operatorname{sech}^2(E/2T)}{32TE^4}\\
			&-\frac{\operatorname{sech}^2(E/2T)\tanh(E/2T)}{32T^2E^3}.
		\end{split}
		\label{eq:frg-sum-S3}
	\end{align}
	Their zero-temperature limits are obtained continuously, giving $S_2(E,0)=1/(4E^3)$ and $S_3(E,0)=3/(16E^5)$.
	With $T=k\tau$ and $m=k\tilde m$, the dimensionless threshold functions appearing in Eq.~\eqref{eq:frg-beta-functions} consequently are given by
	\begin{align}
		\widehat{\mathcal M}(\tau,\tilde m)&=\frac{k^3}{6\pi^2}S_2(E,T),
		\label{eq:frg-threshold-M}\\*
		\widehat{\mathcal A}(\tau,\tilde m)&=\frac{k^3}{6\pi^2}
		[S_2(E,T)-2m^2S_3(E,T)],
		\label{eq:frg-threshold-A}\\*
		\widehat{\mathcal B}(\tau,\tilde m)&=\frac{k^5}{6\pi^2}S_3(E,T).
		\label{eq:frg-threshold-B}
	\end{align}
	The powers of $k$ cancel those in $E$ and $T$, so these functions depend only on $\tau$ and $\tilde m$. At $T=m=0$, they give $\widehat{\mathcal M}=\widehat{\mathcal A}=1/(24\pi^2)$ and $\widehat{\mathcal B}=1/(32\pi^2)$, reproducing the vacuum normalization used in the fixed-point analysis. For $T\gg E$, the sums are suppressed and the fermionic contributions to the flow become small.

	\begingroup
	\subsection{Numerical determination of the pole scale}
	\label{app:frg-pole-numerics}

	For Figs.~\ref{fig:frg-vacuum-mass-benchmark}--\ref{fig:frg-thermal-pole-scale}, we evolve the inverse couplings $e^{2t}/(\tilde G_S\pm\tilde C_4)$ together with the mass. As $t$ decreases from zero, the first zero of $e^{2t}/(\tilde G_S-\tilde C_4)$ gives the pole time. To estimate $T_c$, we find the temperature at which this inverse coupling reaches zero at the end of the integration, $t=-18$. We use Brent's method with an initial bracket of $0.30\leq T/\Lambda\leq0.50$.

	The integration uses machine precision, with accuracy and precision goals of $14$ and $12$, respectively. The $T=T_c$ curve in Fig.~\ref{fig:frg-generated-thermal-flows} uses $T/\Lambda=0.33762662362215523$ before rounding. For this curve, we subtract the small value of the inverse coupling at $t=-18$ when converting back to the plotted couplings.

	At positive temperature, the formulas for $S_2$ and $S_3$ contain terms that nearly cancel when $E/(2T)$ is small. We therefore use their series expansions for $E/(2T)<10^{-3}$.
	\endgroup
	\endgroup
	
	\twocolumngrid
	\FloatBarrier
	\allowdisplaybreaks

	\bibliography{references}
\end{document}